\documentclass[prb]{revtex4}%
\usepackage{graphicx}

\begin{document}
\author
{Alexandre M. Zagoskin}
\affiliation{Physics and Astronomy Dept., The University of British Columbia,
6224 Agricultural Rd., Vancouver, B.C., V6T 1Z1, Canada.\footnote{Formerly also at: D-Wave Systems Inc., 320-1985 W Broadway, Vancouver, B.C., V6J 4Y3, Canada.\\
$^{\dagger}$Lectures at the summer school "Quantum Computation at the Atomic Scale", Istanbul, 2003; Turk. J. Phys. {\bf 27} (2003) 491; references updated.}}

\title{Mesoscopic d-Wave Qubits: Can High-T$_c$ Cuprates  Play  a Role in Quantum Computing?$^{\dagger}$}


\begin{abstract}
 Due to nontrivial orbital pairing symmetry, surfaces and interfaces of high-T$_c$ superconductors
support states which violate time-reversal ($\cal T$-) symmetry.
Such naturally degenerate states, useful as working states of a
qubit, are standard for atomic or molecular-size qubit prototypes
(e.g. based on nuclear spins), but exceptional for mesoscopic
qubits. (In particular, they hold promise of a better
scalability.) In these lectures I review the physics of $\cal
T$-breaking on surfaces and interfaces of high-T$_c$
superconductors; then describe existing proposals for high-T$_c$
based qubits and the current state of experiments; finally, I
discuss the decoherence sources in the system, open questions, and
future research directions. 
\end{abstract}

\maketitle

\section{
Time-reversal symmetry breaking on surfaces and interfaces of high-T$_c$ superconductors}


\subsection{Description of transport in high-T$_c$ structures.}
 The lack of accepted microscopic theory of
superconductivity in high-T$_c$ cuprates did not prevent
successful research in this field since 1986. We have in several
respects the repetition of the situation ca 1938, but with a clear
advantage of already having BCS theory to provide insight and
language for phenomenological treatment.

On this level, high-T$_c$ superconductors are successfully
described by Gor'kov equations for normal and anomalous Green's
functions\cite{Mineev99}, which in Matsubara representation are
defined in the usual way:
\begin{eqnarray} G_{\alpha\beta}({\bf
k},\tau;{\bf k}',\tau')=-\left<{\cal T}_{\tau}a_{\alpha}(\tau)
a^{\dagger}_{\beta}(\tau')\right>,\nonumber\\
F_{\alpha\beta} =({\bf k},\tau;{\bf k}',\tau')=\left<{\cal
T}_{\tau}a_{{\bf k}\alpha}(\tau) a^{\dagger}_{-{\bf
k}'\beta}(\tau')\right>,\\
F_{\alpha\beta}^+ =({\bf k},\tau;{\bf k}',\tau')=\left<{\cal
T}_{\tau}a_{-{\bf k}\alpha}(\tau) a^{\dagger}_{{\bf
k}'\beta}(\tau')\right>.\nonumber
\end{eqnarray}

The only difference from conventional superconductivity is in the
nontrivial symmetry of the pairing potential,
\begin{equation}
H_{\rm int} = \frac{1}{2}\sum_{{\bf k},{\bf k}',{\bf
q}}V_{\alpha\beta,\lambda\mu}({\bf k},{\bf k}')a^{\dagger}_{-{\bf
k}+{\bf q}/2, \alpha}a^{\dagger}_{{\bf k}+{\bf q}/2,\beta} a_{{\bf
k}'+{\bf q}/2,\lambda}a_{-{\bf k}'+{\bf q}/2,\mu}.
\end{equation}

As usual, we write equations of motion for each of these functions
and use the "anomalous mean field" recipe to decouple the
four-operator products,
$\left<a^{\dagger}a^{\dagger}aa\right>\to\left<a^{\dagger}a^{\dagger}\right>\left<aa\right>$.
This brings out modified self-consistency relations for the order
parameter,
\begin{eqnarray}\label{eq_self-consistency}
\Delta_{\alpha\beta}({\bf k},{\bf q}) = -\sum_{{\bf
k}'}V_{\beta\alpha,\lambda\mu}({\bf k},{\bf k}')
F_{\lambda\mu}({\bf k}'+{\bf q}/2,\tau;{\bf k}'-{\bf
q}/2,\tau),\nonumber\\
 \Delta_{\lambda\mu}^+({\bf k},{\bf q}) =
-\sum_{{\bf k}'}V_{\alpha\beta,\mu\lambda}({\bf k}',{\bf k})
F_{\alpha\beta}^+({\bf k}'-{\bf q}/2,\tau;{\bf k}'+{\bf
q}/2,\tau),
\end{eqnarray}
and Gor'kov equations, which in the spatially uniform, stationary
system read:
\begin{eqnarray}\label{eq_Gorkov}
(i\omega_n-\xi_k) G_{\alpha\beta}({\bf k},\omega_n) +
\Delta_{\alpha\gamma}F_{\gamma\beta}^+({\bf k},\omega_n) = \delta_{\alpha\beta} ;\nonumber\\
(i\omega_n+\xi_k) F_{\alpha\beta}^+({\bf k},\omega_n) +
\Delta_{\alpha\gamma}^+G_{\gamma\beta}({\bf k},\omega_n) = 0;\\
(i\omega_n-\xi_k) F_{\alpha\beta}({\bf k},\omega_n) -
\Delta_{\alpha\gamma}G_{\beta\gamma}(-{\bf k},-\omega_n) =
0.\nonumber
\end{eqnarray}
Here $\xi_k$ is the fourier transform of the kinetic energy
operator, $\hat{\xi}=(2m)^{-1}\hat{p}^2-\mu$, $\mu$ being the
chemical potential.

It was established, that in high-T$_c$ cuprates, like YBCO, the
order parameter is a spin singlet with $d$-wave orbital symmetry,
\begin{equation}\label{eq_d_wave_symmetry}
\Delta_{\alpha\beta}({\bf k})=\delta_{\alpha\beta}\Delta({\bf
k}),\:\:\Delta({\bf k})\propto \cos^2(k_x)-\cos^2(k_y)
\end{equation}
(with axes chosen along crystallographic directions (1,0,0),
(0,1,0) in the cuprate
layer)\cite{Sigrist95,VanHarlingen95,Tsuei00}. For the following,
the most important consequence of this symmetry is the sign change
of $\Delta({\bf k})$ for certain directions. This means, first,
that there exists an {\em intrinsic phase shift} of $\pi$ between
different directions in the crystal; second, that in certain
(nodal) directions the order parameter is zero, and therefore the
quasiparticle excitation spectrum is not gapped.

On the spatial scale exceeding the coherence length, $\xi_0$, it
is more convenient to use the Eilenberger
equations\cite{Eilenberger68}, which follow from (\ref{eq_Gorkov})
in the quasiclassical limit. This is certainly justified in
high-T$_c$ cuprates with their small $\xi_0$.

The Eilenberger equations are conveniently written in matrix form,
\begin{eqnarray}\label{eq_Eilenberger}
{\bf v}_F\cdot\hat{G}(\omega_n)+[\omega_n
\hat{\tau}_3+\hat{\Delta},{\hat G}(\omega_n)]=0.
\end{eqnarray}
Here the matrix Green's function and order parameter,
\begin{eqnarray}\label{eq_Eilenberger_definitions}
\hat{G}({\bf v}_F,{\bf r};\omega_n) =\left(\begin{array}{cc}
  g_{\omega_n}({\bf v}_F,{\bf r}) & f_{\omega_n}({\bf v}_F,{\bf r}) \\
  f_{\omega_n}^+({\bf v}_F,{\bf r}) & g_{\omega_n}({\bf v}_F,{\bf
  r})\end{array}\right); \:\:\:\hat{\Delta}({\bf v}_F,{\bf r};\omega_n) =\left(\begin{array}{cc}
0 & \Delta({\bf v}_F,{\bf r}) \\
  \Delta^+({\bf v}_F,{\bf r}) & 0
\end{array}\right),
\end{eqnarray}
depend both on position $\bf r$ and on (direction of) Fermi
velocity ${\bf v}_F$. (The layered structure of high-T$_c$
cuprates allows us to reduce the problem to two dimensions to a
good accuracy, unless we have to consider, for example, a twist
junction\cite{Tafuri00}, or tunneling in the $c$-direction.) The
components of $\hat{G}$, obtained from Gor'kov's functions by
integration over energies, satisfy the normalization condition,
$g_{\omega_n}=\sqrt{1-f_{\omega_n}^+f_{\omega_n}},$ and the
self-consistency relation (\ref{eq_self-consistency}) becomes
\begin{equation}\label{eq_self-consistency_bis}
\Delta({\bf v}_F,{\bf r}) = 2\pi
N(0)T\sum_{\omega_n>0}\left<V({\bf v}_F,{\bf
v}_F')f_{\omega_n}({\bf v}_F,{\bf r}) \right>_{\theta}.
\end{equation}
Here the angle averaging $\langle\rangle_{\theta} =
\int_0^{2\pi}\frac{d\theta}{2\pi}$.

 In a little different language, the same
results are obtained with the Andreev approximation in the
Bogoliubov-de Gennes equations for the components of the
single-bogolon wave function, $(u({\bf r}), v({\bf r}))^{\top}$.
The original Bogoliubov-de Gennes equations are obtained in the
process of diagonalization of pairing BCS Hamiltonian
\cite{Zagoskin98}:
\begin{eqnarray}\label{eq_BdG}
\left(\begin{array}{cc}-\frac{1}{2m}\nabla^2-\mu & \Delta_k({\bf
r})\\
\Delta_k^*({\bf r}) & \frac{1}{2m}\nabla^2 + \mu\end{array}
\right)\left(\begin{array}{c}u_k({\bf
r})\\
v_k({\bf r})\end{array}\right) \approx
\left(\begin{array}{cc}-{\bf v}_F\cdot\nabla & \Delta_k({\bf
r})\\
\Delta_k^*({\bf r}) & {\bf v}_F\cdot\nabla\end{array}
\right)\left(\begin{array}{c}u_k({\bf
r})\\
v_k({\bf r})\end{array}\right) = E_k\left(\begin{array}{c}u_k({\bf
r})\\
v_k({\bf r})\end{array}\right).
\end{eqnarray}
Here the quasimomentum $\bf k$ labels the bogolon state, $E_k$ is
the excitation energy, and the self-consistency relation reads
\begin{equation}\label{eq_self-consistency_BdG}
\Delta^*_k({\bf r},T)=\sum_{k'}V({\bf k},{\bf k}')u_{k'}^*({\bf
r})v_{k'}({\bf r}) \tanh \frac{E_{k'}[\Delta_{k'}({\bf r})]}{2T}.
\end{equation}

Unlike the Gor'kov equations (\ref{eq_Gorkov}) (or the initial
Bogoliubov-de Gennes equations), the equations
(\ref{eq_Eilenberger},\ref{eq_BdG}) are of the first order in
gradients, which allows us to introduce quasiclassical
trajectories (characteristics) along ${\bf v}_F$ and solve the
corresponding equations by integration along these trajectories,
with proper boundary conditions.

  It is known that a
quasiparticle (electron or hole, described, in terms of
Bogoliubov-de Gennes equations, by a vector
$\vec{\psi}_k=(1,0)^{\top}\exp(i{\bf k r})$
($(0,1)^{\top}\exp(i{\bf k r})$) respectively) impinging on the
superconductor from the normal metal can undergo an {\em Andreev
reflection}, switching the branch of the excitation spectrum,
acquiring an additional phase, and almost exactly reversing the
direction of its group velocity (this happens because for an
electron and a hole with the same momentum $\bf k$ group
velocities are opposite):
\begin{equation}\vec{\psi}\to \hat{\cal
R}_A\cdot\vec{\psi},\end{equation} where
\begin{eqnarray}\label{eq_Andreev_bc}
\hat{\cal R}_A=\left(\begin{array}{cc} 0 & e^{-i\pi/2+i\chi}\\
e^{-i\pi/2-i\chi} & 0\end{array} \right).
\end{eqnarray}
 The phase  $\chi$ is the
 phase of the superconducting order parameter; the ($-\pi/2$)-shift
 is exact in the limit when the quasiparticle energy is much less
 than the superconducting gap (generally it is some energy-dependent
 function $\delta(E)$).

Now consider a slab of normal conductor sandwiched between two
superconductors ({\em SNS junction}). If we neglect the spatial
dependence of the order parameter in superconductors, we don't
need to solve the self-consistency equations
(\ref{eq_self-consistency},\ref{eq_self-consistency_BdG}).
Therefore the problem reduces to a single-particle one and is most
naturally solved in Bogoliubov-de Gennes language
(Eq.(\ref{eq_BdG}) becomes a Schr\"{o}dinger equation for a
two-component wave function).

Solutions of this equation with the boundary condition
(\ref{eq_Andreev_bc}) are standing waves. Obviously in the act of
Andreev reflection a charge of $\pm 2e$ is transferred to the
superconductor, therefore every standing wave (Andreev level)
carries supercurrent.

Quasiclassically, in order to find Andreev levels in a normal
layer of thickness $L$, sandwiched between "left" and "right"
superconductors, with phase difference $\chi$, we write the
Bohr-Sommerfeld quantization condition,
\begin{equation}\label{eq_Bohr_Sommerfeld}
\oint p(E)dq \pm \chi + \delta_l(E)+\delta_r(E) = 2\pi n.
\end{equation}
Here the kinematic phase gain of the quasiparticle along the
closed trajectory, $\oint p(E)dq = \int_l^r p_e(E) dq + \int_r^l
p_h(E) dq = \int_l^r (p_e(E)-p_h(E)) dq$, takes into account
electron-hole (or vice versa) conversion.

The positions of levels, and therefore the supercurrent, depend on
the phase difference $\chi$ between the superconducting banks, and
we arrive at Josephson effect in SNS structures
\cite{Kulik70,Ishii70,Bardeen72}. Actually, the language of
Andreev levels can be successfully used to describe the Josephson
effect in general (for a review see \cite{Furusaki99}).

\subsection{$\pi$-junctions and time-reversal symmetry
breaking}\label{sec_1_pi_junction}

The crucial experiments(\cite{Wollman93,Wollman95,Tsuei94}; see
also review\cite{Tsuei00}) which confirmed $d$-wave pairing
symmetry in high-T$_c$ cuprates were directed at catching the
intrinsic phase $\pi$-shift. The general idea of the experiment
follows from the fluxoid quantization condition in a
superconductor: if a superconducting contour $\cal C$ is
penetrated by the magnetic flux $\Phi$, then \cite{Tinkham96}
\begin{equation}\label{eq_fluxoid_quantization}
2\pi\frac{\Phi}{\Phi_0}+\oint_{\cal C}d{\bf s}\cdot\nabla\phi =
2\pi n,\:\:\Phi_0\equiv \frac{hc}{2e}
\end{equation}
(in CGS units; $\Phi_0 \approx 2\cdot 10^{-15}$Wb in SI).

In the case of a massive superconducting ring the contour can be
chosen well inside the superconductor, where there is no current
and therefore no superconducting phase gradient. Then the magnetic
flux is quantized in  units of $\Phi_0$.

 If there is a Josephson junction in the ring, the phase change
 will concentrate there, yielding
\begin{equation}\label{eq_fluxoid_quantization_2}
2\pi\frac{\Phi}{\Phi_0}+\chi = 2\pi n.
\end{equation}
Here  $\chi$ is the phase difference across the junction.

The equilibrium value of $\chi$ is determined by the interplay
between Josephson and magnetic energy of the system. The Josephson
energy is related to the Josephson current via
\begin{equation}\label{eq_Josephson_relation}
I(\chi)=2e\frac{\partial E(\chi)}{\partial\chi},
\end{equation}
and in the simplest case of a tunneling junction,
$I(\chi)=I_c\sin\chi$, and $E(\chi)=-(I_c/2e)\cos\chi.$ For a
conventional Josephson junction $E(\chi)$ is at a minimum when
$\chi=0$; therefore in the absence of an external field the total
energy of the ring,
\begin{eqnarray}\label{eq_ring_energy}
U(\chi)=E(\chi)+E_{\rm magn.}(\Phi)=-\frac{I_c}{2e}\cos\chi +
\left(\frac{\chi}{2\pi}\right)^2\frac{\Phi_0^2}{2L},
\end{eqnarray}
has a single  global minimum at $\chi=0$ for any value of $2\pi
LI_c/\Phi_0$ (here $L$ is the self-inductance of the ring).

Not so if the pairing symmetry is $d$-wave: if we choose the
configuration of the loop in such a way, that the opposite sign
lobes contact across the junction (so called $\pi$-junction), the
current-phase relation switches to
$I(\chi)=I_c\sin(\chi+\pi)=-I_c\sin\chi$, and
$E(\chi)=+(I_c/2e)\cos\chi.$ Therefore if $2\pi LI_c/\Phi_0>1,$
the system has two degenerate minima; if this ratio is so big that
the magnetic energy can be neglected compared to the Josephson
energy, the equilibrium phase difference is $\chi_0=\pi$, and we
obtain from (\ref{eq_fluxoid_quantization_2}), that
\begin{equation}\label{eq_fluxoid_quantization_3}
\Phi = \left(n+\frac{1}{2}\right)\Phi_0.
\end{equation}
Thus in equilibrium there is a spontaneous flux $\Phi_0/2$ in the
ring. Its direction (up or down) allows us to distinguish the two
ground states of the system, where the \emph{time-reversal
symmetry is thus broken.}

Such behaviour was indeed observed in an experiment
\cite{Tsuei94}: a tri-crystal ring of YBCO generated a half-flux
quantum, which was detected by SQUID microscopy.

A natural question is whether $\pi$-junctions are the only
possibility provided by $d$-wave symmetry, and whether only
$\Phi_0/2$ fluxes can be spontaneously generated. The answer is
no: in principle, any equilibrium phase difference can be realized
in a $d$-wave junction, and time-reversal symmetry breaking can be
accompanied by generation of an arbitrary magnetic flux, or none
at all.

\subsection{Josephson effect and {\cal T}-breaking in SND and DND
junctions}\label{sec_1_Josephson effect}

Let us return to an SNS junction (assuming a rectangular normal
part, $L\times W$). Each quasiclassical trajectory connecting two
superconductors acts as a conduit of supercurrent between them, an
"Andreev tube" of diameter $\sim\lambda_F$, which carries
supercurrent, determined by the phase difference between its ends.
If we neglect the normal scattering at NS interfaces, there is no
mixing between different trajectories, since Andreev reflections
simply reverse the velocity, sending the reflected particle along
the same path.
It can be shown, that if $L\gg\xi_0$, every such
trajectory passing through the point $\bf r$ contributes a partial
supercurrent density\cite{Zagoskin98,Barzykin99}
\begin{equation}\label{eq_partial_current_density}
{\bf j}({\bf r},{\bf v}_F) = \frac{2e{\bf v}_F}{\lambda_F
W}\sum_{n=1}^{\infty} (-1)^{n+1}\frac{{\cal L}({\bf
v}_F)}{l_T}\frac{\sin (n(\phi_1-\phi_2))}{\sinh (n{\cal L}({\bf
v}_F)/l_T)}e^{-2n{\cal L}({\bf v}_F)/v_F\tau_i}.
\end{equation}

In Eq.(\ref{eq_partial_current_density}), $\cal L$(${\bf v}_F$) is
the length of the trajectory, and $l_T=v_F/2\pi k_BT$ is the so
called normal metal coherence length. (We also included effects of
weak impurity scattering with scattering time $\tau_i$.) The
physical meaning of $l_T$ is that in a clean normal metal an
electron and Andreev-reflected hole (or vice versa) with energy
$k_BT$ maintain phase coherence across distance $l_T$ simply
because they travel along the same trajectory. Indeed, the
momentum of an electron (hole) with energy $k_BT\ll E_F$ is
$p_{e,h}\approx p_F\pm (\partial_E p_F)k_BT = p_F \pm k_BT/v_F.$
At a distance $l$ from the point of Andreev reflection (measured
along the trajectory) they would gain phase difference
$(p_e-p_h)l=2k_BTl/v_F.$ If the phase difference is of order
$\pi$, the coherence is effectively lost, so we get $l_{T,{\rm
ballistic}} \sim v_F/k_BT$. The factor of $2\pi$ appears in
accurate
 treatment, like in Eq.(\ref{eq_partial_current_density}).

 As an aside, in case of very strong
 scattering in the normal part of the system, when the motion of electrons/holes is diffusive
  with diffusion coefficient $D$, we can still use the same argument. Now the observable length scale
  is given by the displacement of quasiparticle, $l^2=Dt,$ while the phase difference between
  the electron and the hole is gained along the crooked path of length $l'=v_Ft=v_Fl^2/D$ they take.
  So the condition $2k_BTl'/v_F\sim \pi$ yields $l_{T,{\rm diffusive}}\sim\sqrt{D/k_BT}.$

Such coherence in a normal metal is a purely {\em kinematic}
effect, since there is no interaction in the normal (non-magnetic)
metal, which would either support or suppress superconductivity.
("Normal metal is neutral with respect to superconducting
correlations."-C.W.J. Beenakker.)  Nevertheless its effects are
quite real, e.g. the very possibility of coherent supercurrent
flow (Josephson effect) in SNS junctions.

Let us apply the approach of (\ref{eq_partial_current_density}) to
calculation of Josephson current in a clean SNS contact. In every
point of NS boundary (e.g. $x=L$), we integrate
(\ref{eq_partial_current_density}) over directions of ${\bf v}_F$
(such that $v_{Fx}>0$). As it should be expected, the result
reproduces Ishii's sawtooth \cite{Ishii70}: in the limit of zero
temperature and no scattering inside the normal layer
($\chi\equiv\phi_1-\phi_2$),
\begin{equation}\label{eq_Ishii}
I(\chi) = \int dy \sum_{v_{Fx}>0}j_x(y,{\bf v}_F) \propto
\frac{2}{\pi}\sum_{n=1}^{\infty}(-1)^{n+1}\frac{\sin(n\chi)}{n} =
\frac{\chi}{\pi},\:\:|\chi|<\pi
\end{equation}
(periodically extended).


\begin{figure}[tbp]
\begin{center}
\includegraphics[scale=0.8]{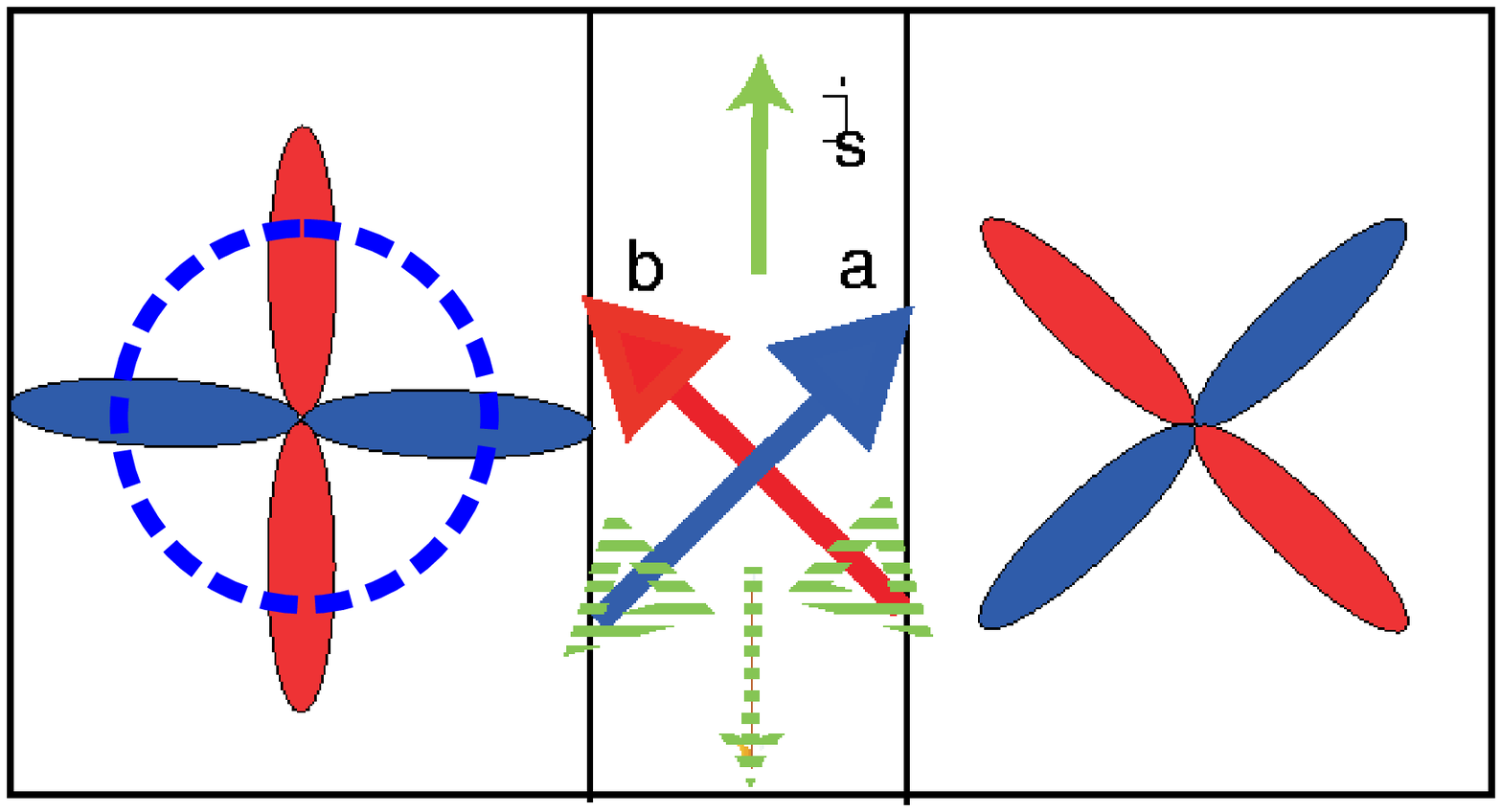}
\end{center}
\caption{Josephson and spontaneous currents in SND and DND
junctions. Arrows $a$ and $b$ indicate Andreev zero- and
$\pi$-levels. In equilibrium their contributions to the Josephson
current (normal to the NS boundary) cancel. By switching current
directions (dashed arrows) we come to the other ground state, with
the opposite direction of the spontaneous current, $j_S$.}
\label{fig_DND}
\end{figure}

\begin{figure}[tbp]
\begin{center}
\includegraphics[scale=0.5]{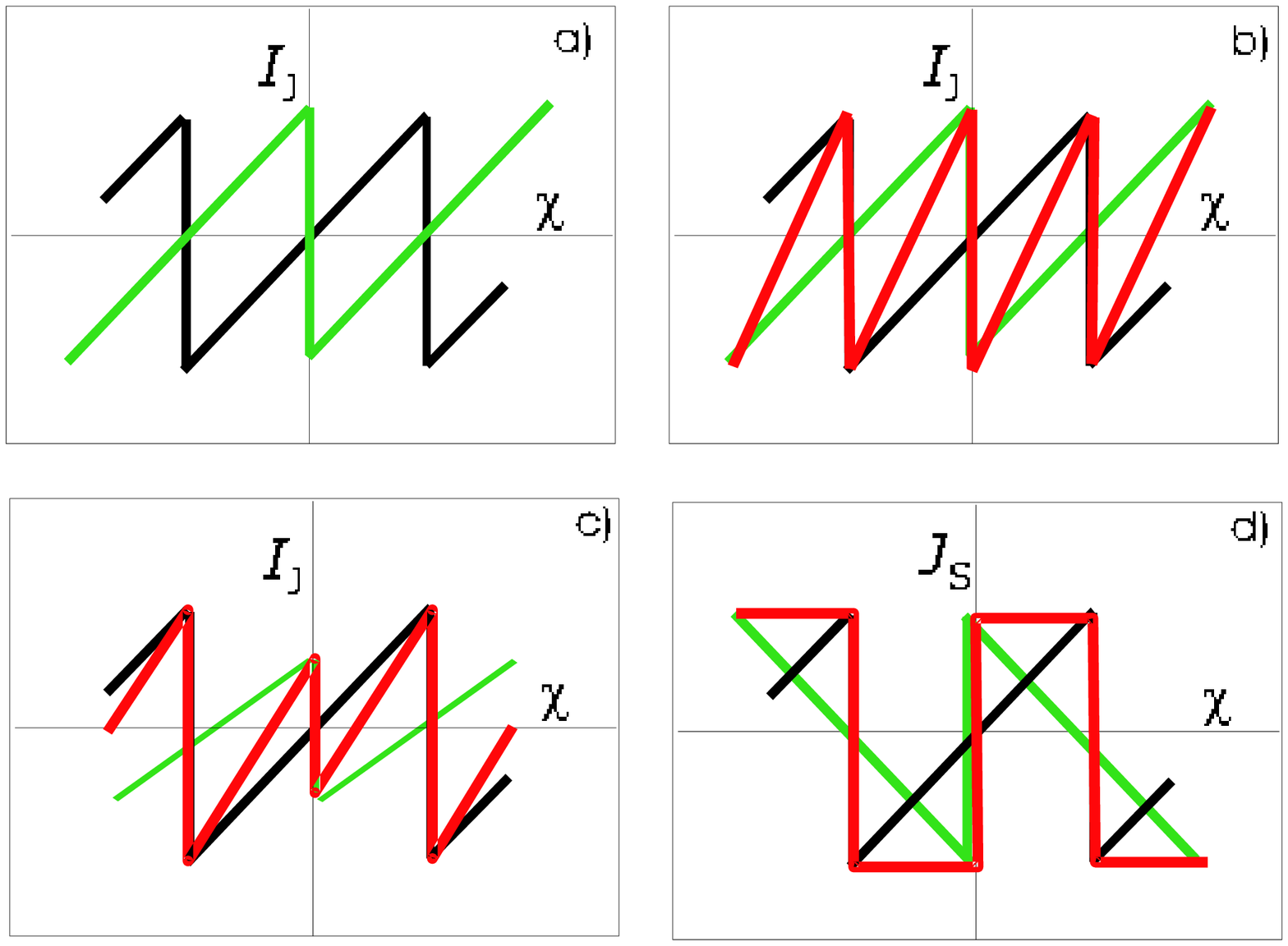}
\end{center}
\caption{Josephson and spontaneous currents in SND and DND
junctions. a)Contributions of zero- and $\pi$-levels to the
Josephson current in a 0-45$^{\circ}$ DD(SD) junction. b)Resulting
current-phase dependence. c) Same in more general case. d)
Spontaneous current in a 0-45$^{\circ}$ junction. }
\label{fig_sum_sawtooth}
\end{figure}

 Consider now an SND junction, that is, an
SNS contact, where one of the superconductors has $d$-wave
symmetry \cite{Zagoskin97,Zagoskin98}. Now, when integrating over
the directions of ${\bf v}_F$ we encounter two kinds of
trajectories: "zero"- and "$\pi$"-trajectories, which link the
conventional superconductor with the lobes of the $d$-wave order
parameter of opposite sign (intrinsic phase $\pi$-shift). The
Josephson current is therefore a sum of two contributions like
(\ref{eq_Ishii}), one of which has the phase argument $\chi +\pi$.
The relative weight of these contributions depends on the
orientation of the SD boundary with respect to the crystal axes of
the cuprate (to which the order parameter is nailed). In the case
of a $45^{\circ}$ orientation, when the boundary is along the
(110) plane and is therefore normal to the nodal direction of the
$d$-wave order parameter, the two groups contribute equally,
yielding ($\cos\theta=v_{Fx}/v_F$)
\begin{equation}\label{eq_Ishii_d_wave}
I(\chi)=\sum_{v_{Fx}>0}^{\rm zero\:
levels}\frac{ev_{F}\cos\theta}{L}\frac{2}{\pi}\sum_{n=1}^{\infty}
(-1)^{n+1}e^{-2nL/l_i\cos\theta}\frac{L}{l_T\cos\theta}\frac{\sin(n\chi)+\sin(n(\chi+\pi))}
{\sinh(nL/l_T\cos\theta)}.
\end{equation}
All odd harmonics cancel, and we obtain a $\pi$-periodic sawtooth:
at $T=0$, $l_i\to\infty$
\begin{equation}\label{eq_Ishii_d_wave_2}
I(\chi)\propto \frac{\chi-\pi/2}{\pi/2}, \:\:0<\chi<\pi,
\end{equation}
(periodically extended). (This is simply a sum of two identical,
$2\pi$-periodic sawtooth functions, shifted by $\pi$, see
Fig.\ref{fig_sum_sawtooth}.) There are two stable equilibrium
phase differences across the junction: $\chi=\pm\pi/2.$ This
means, that the time-reversal symmetry is broken. (In addition,
the frequency of ac Josephson effect in the system will be doubled
\cite{Zagoskin97,Hurd99}).

 The SND junction may seem simply a "superposition" of two SNS junctions, with phase differences
 shifted by $\pi$, but the situation is more interesting. In equilibrium, spontaneous currents flow in the normal layer,
parallel to the boundary\cite{Huck97}. This is clear from our
"Andreev tube" treatment. Take, for example, two trajectories,
with $\theta=+\alpha$ and $\theta=-\alpha$, carrying equal partial
currents. Their contribution to the Josephson current (normal to
the boundary) is zero, since their currents' projections on this
direction cancel each other. On the contrary, the projections on
the direction, parallel to the boundary, add. The phase dependence
of this current, $I_s(\chi)$, can be obtained in the same way as
for the Josephson current. In the limit of zero temperature and
clean normal layer it reduces to a {\em difference} of two
sawtooth functions, yielding
\begin{equation}\label{eq_spontaneous_current_SND}
I_s(\chi)\propto {\rm sgn}\chi,\:\:|\chi|<\pi,
\end{equation}
again periodically extended. States with $\pi/2$ and $-\pi/2$ thus
carry spontaneous currents in opposite directions. If an SND
junction is closed on itself (annular geometry), these currents
translate into spontaneous magnetic fluxes, normal to the plane of
the system \cite{Zagoskin98a}.

Note that all of the above is only possible if higher harmonics of
current-phase dependence are not all zero, since all the odd
harmonics (including the standard Josephson term, $\sin\chi$)
cancel. (The expected "total depairing" in the (110) plane.)

In the case of arbitrary orientation of the ND plane , the weights
of zero- and $\pi$-sawtooth functions will be different
(Fig.\ref{fig_sum_sawtooth}c). Then the current-phase dependence
regains $2\pi$-periodicity. At zero temperature and in the absence
of scattering, the {\cal T}-breaking is generally still present.
The equilibrium phase difference is no longer $\pm\pi/2$ and
depends on the geometry. (The exception is when the boundary is
along (100) or (010). Then the contribution of only one group
survives, and we have either standard, or $\pi$-junction. Of
course, it is impossible to tell, whether a single junction is a
$\pi$-junction - it is necessary to look at the contour, in which
it is included\cite{VanHarlingen95}.) Finite temperature and
scattering suppress higher harmonics first, and therefore the
transition to non-{\cal T}-breaking state is possible
\cite{Huck97}.

\begin{figure}[tbp]
\begin{center}
\includegraphics[scale=0.8]{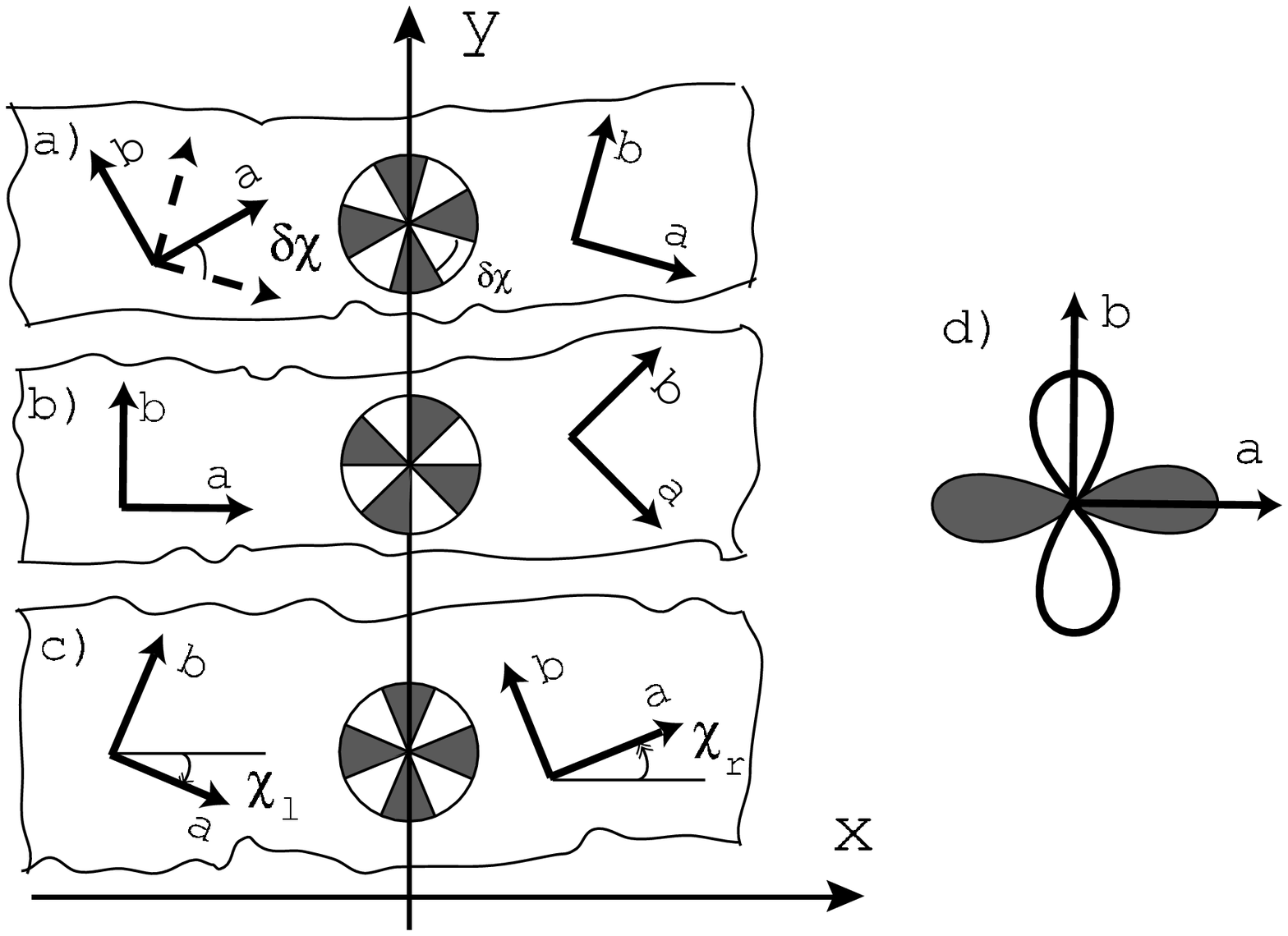}
\end{center}
\caption{Josephson and spontaneous currents in DND junctions
\cite{Amin01,Amin02b}. a) Arbitrary case. b) Asymmetric
(0-45$^{\circ}$) junction. c) Symmetric
(22.5$^{\circ}$-22.5$^{\circ}$) junction. Directions corresponding
to zero-levels are shaded. In the case c) the contributions of
zero- and $\pi$-levels to the current in $y$-direction are
identically zero; therefore the spontaneous current is absent, but
the time-reversal symmetry is still broken. d) Orientation of the
$d$-wave order parameter with respect to the crystalline axes. }
\label{fig_DDcurrents}
\end{figure}

Essentially the same interplay of zero- and $\pi$-levels takes
place in DND junctions. There we have a somewhat richer picture.
For example, the time-reversal symmetry can be broken without
producing spontaneous currents (Fig.\ref{fig_DDcurrents}c).

\begin{figure}[tbp]
\begin{center}
\includegraphics[scale=0.75]{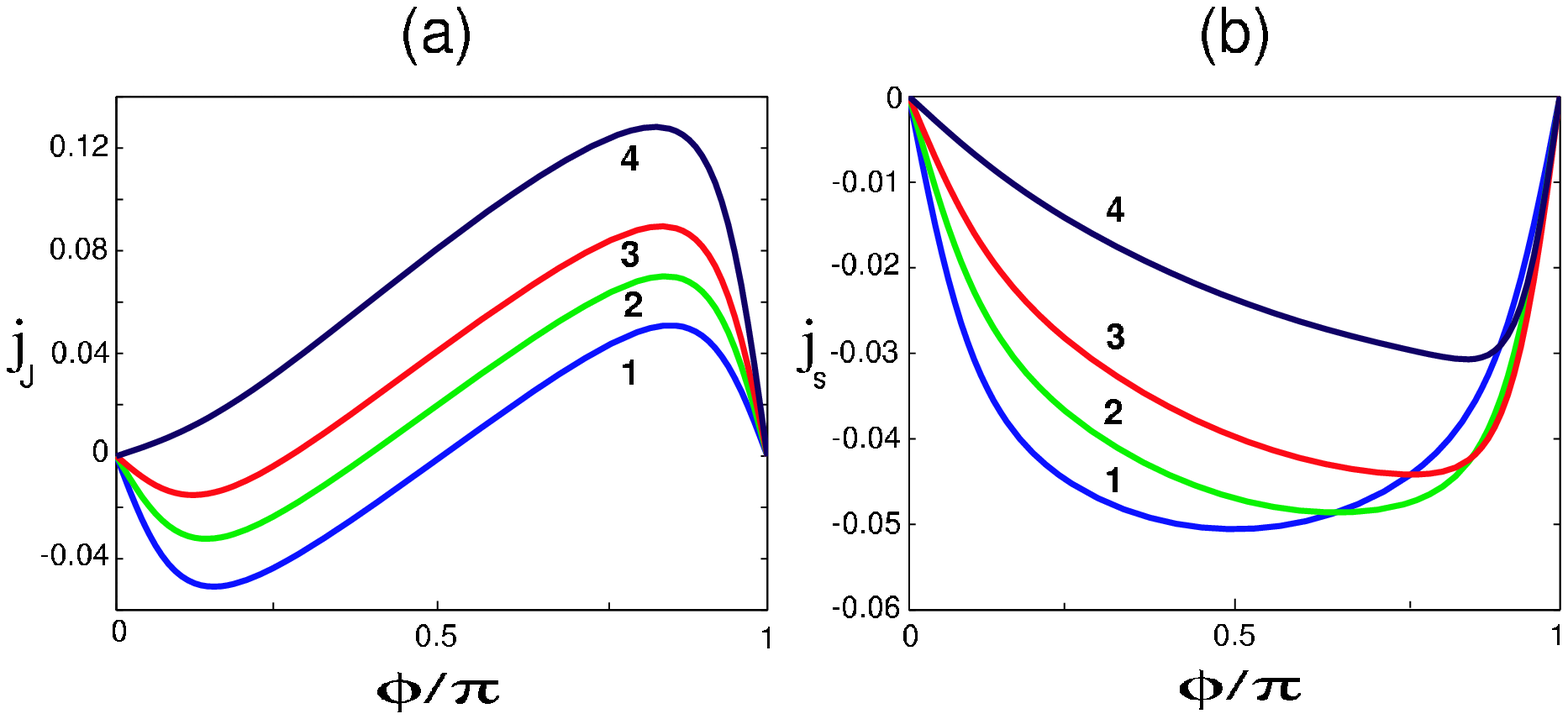}
\end{center}
\caption{Josephson current (a) and spontaneous current (b) versus
the phase difference in a clean DD grain boundary junction
calculated in the non-self-consistent approximation. Temperature
is $T=0.1T_{c}$. The mismatch angles are $\chi_l=0$ and $\chi_r =
45^\circ$ (1),\ $40^\circ$ (2),\ $34^\circ$ (3),\ $22.5^\circ$ (4)
\cite{Amin02b}. }\label{fig_spontaneous current phase}
\end{figure}

\begin{figure}[tbp]
\begin{center}
\includegraphics[scale=0.75]{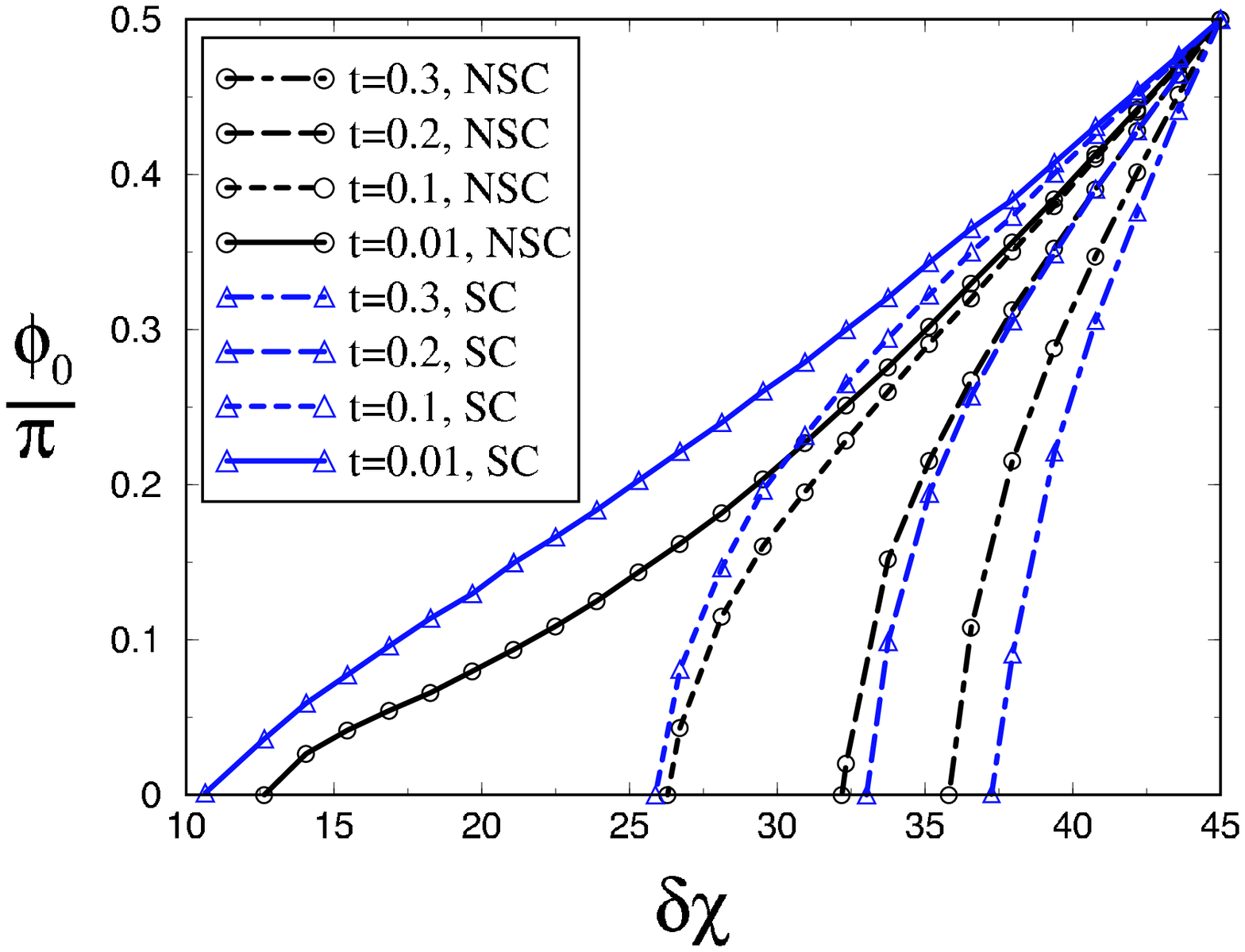}
\end{center}
\caption{Equilibrium phase difference $\phi_0$ in a clean grain
boundary junction as a function of the mismatch angle: $\chi_l=0$;
$\chi_r=\delta\chi$ (Fig. \ref{fig_DDcurrents}). Parameter
$t=T/T_c$ is the dimensionless temperature. Triangles and circles
correspond to self-consistent and non-self-consistent solutions of
Eilenberger equations, respectively. The 0-45$^{\circ}$ junction
is  a $\pi$/2-junction at any temperature, since by symmetry all
odd components of the Josephson current are cancelled. Compare
this to the behaviour observed in YBCO junctions
(Fig.\ref{fig_Ilichev01}).
 }\label{fig_bistability_region}
\end{figure}

One of the reliable methods of fabrication of Josephson devices in
high-T$_c$ cuprates is based on forming grain boundary junctions
(see e.g. \cite{Tafuri99}). The order parameter is suppressed
within $\sim\xi_0$ around the boundary, and this region can be
considered "normal". Therefore the SND/DND model applies, but only
qualitatively, since the above equations are derived in the limit
$L\gg\xi_0$. A more accurate approach, based on Eilenberger
equations (\ref{eq_Eilenberger}), confirms the qualitative
similarity between DND and DD junctions\cite{Amin01}.

\begin{figure}[tbp]
\begin{center}
\includegraphics[scale=0.8]{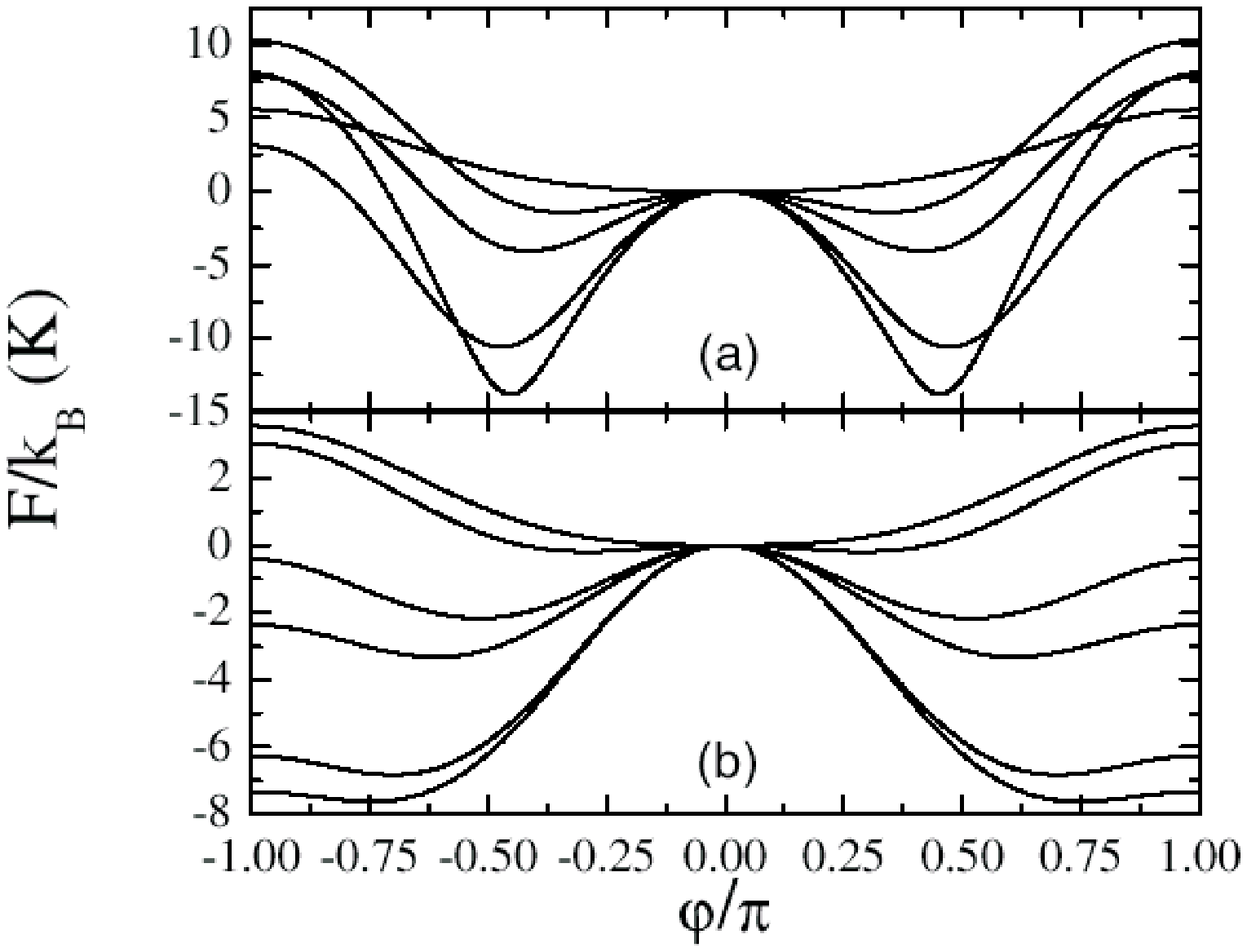}
\end{center}
\caption{Bistability in a grain boundary YBCO
junction\cite{Ilichev01}. Free energy, restored from Josephson
current-phase dependence, is plotted vs. phase for (a)
0-45$^{\circ}$  and (b) symmetric (22.5$^{\circ}$-22.5$^{\circ}$)
junction. Curves (top to bottom) correspond to temperatures (a)
30, 20, 15, 10, and 4.2 K; (b) 20, 15, 11, 10, 5, and 1.6 K,
respectively. } \label{fig_Ilichev01}
\end{figure}

Integrating the current-phase dependences of
Fig.\ref{fig_sum_sawtooth}, we obtain the Josephson energy of the
junction, Eq.(\ref{eq_Josephson_relation}), which has two minima,
corresponding to its degenerate ground states. This bistability
plays the crucial role in qubit applications of high-T$_c$
superconductors. This prediction was confirmed in grain boundary
YBCO junctions\cite{Ilichev01} (Fig.\ref{fig_Ilichev01}).

\begin{figure}[tbp]
\begin{center}
\includegraphics[scale=0.5]{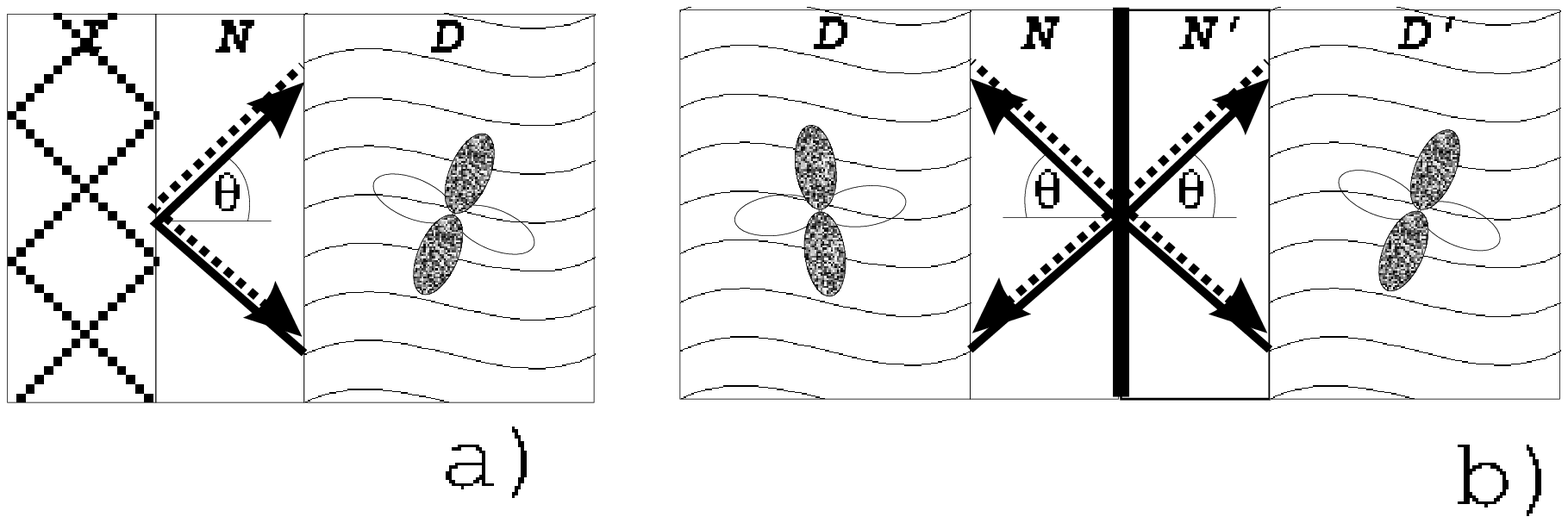}
\end{center}
\caption{Surface bound state in a $d$-wave superconductor (a) and
model of a DD junction with finite transparency (b) (after
\cite{Lofwander}) .} \label{fig_INS}
\end{figure}

So far we did not take into account normal scattering on NS
interfaces, and therefore missed an important point. Let us first
consider a boundary of a $d$-wave superconductor with a vacuum or
an insulator. The order parameter near the surface is suppressed,
and a qualitative understanding of the situation can be obtained
by "inserting" a normal layer of thickness $\sim \xi_0$ between
the insulator and the bulk superconductor, similar to the DND
model\cite{Lofwander}. Energy levels in such a layer can be found
from (\ref{eq_Bohr_Sommerfeld}) for every quasiparticle trajectory
(assuming specular reflection), Fig.\ref{fig_INS}a:

\begin{equation}\label{eq_Bohr_Sommerfeld_INS}
\oint p(E)dq +2\delta(E) + \pi s = 2\pi n.
\end{equation}
Here $s=0$ or $s=1$ depending on whether the trajectory connects
the lobes of the same or opposite sign. Note that $\oint p(0)dq =
0$ and $2\delta(0)=\pi$. Therefore solution $E=0$  exists if and
only if $s=1$. This is the zero-energy, or midgap, state (ZES,
MGS), which obviously cannot exist in conventional
superconductors, where $s$ always equalis zero  (see review
\cite{Lofwander}).

Now consider a DD junction with finite transparency, ${\cal D} <
1$ (instead of the case of ideal transparency, ${\cal D}=1$, which
we dealt with earlier). We can use the DND model, inserting in the
middle of the normal layer an infinitely thin barrier with
transparency ${\cal D}(\theta)$, dependent on the incidence angle
of the quasiparticle trajectory\cite{Lofwander}.

Bohr-Sommerfeld quantization conditions for quasiparticle
trajectories shown in Fig.\ref{fig_INS}b yield the energy levels
of the bound states in the junction. In the limit of low
transparency, ${\cal D} \ll 1$, the critical current is much
larger ($O(\sqrt{\cal D})$, not $O({\cal D})$), if the
orientations of the $d$-wave order parameter allow formation of
zero energy states on both sides of the barrier.

Even when transparency is not small, the presence of ZES may be
qualitatively important. For example, the junction can become a
$\pi$-junction at low enough temperature (see review
\cite{Lofwander}); you can see such behaviour in
Fig.\ref{fig_Ilichev01}b, where below 11 K the potential wells are
centered around $\pi$ rather than zero.


The appearance of spontaneous currents in SD and DD junctions can
be considered as due to a time-reversal symmetry breaking order
parameter with $s+e^{i\chi_0}d$ or $d+e^{i\chi_0}d'$ symmetry,
which is formed in the junction area due to the proximity effect
(here $\chi_0 \neq 0,\pi$ is the equilibrium phase difference
across such junction). In certain conditions, such combinations
could appear near the surface of a $d$-wave superconductor,
leading to spontaneous currents and magnetic moments. So far there
is no conclusive evidence for such currents (see \cite{Amin02} and
references therein).

\section{
 Mesosopic $d$-wave qubits}\label{sec_2}

\subsection{Flux qubits with conventional superconductors}

Qubits are the basic building blocks of future quantum computers.
Essentially they are two-state quantum systems which can be put in
an arbitrary superposition of states ("initialized"), coupled to
each other, undergo desired quantum evolution and measured ("read
out") before losing quantum coherence. Here we concentrate on {\em
superconducting phase qubits}.

The simplest example of such a qubit is an RF SQUID, that is, a
loop with a single Josephson junction (like in Section
\ref{sec_1_pi_junction}). We saw, that in certain circumstances
the system has two degenerate minima, corresponding to a flux of
$\pm\Phi_0/2$ through the loop.

The Hamiltonian of such a system is

\begin{equation}\label{eq_Hamiltonian_phase_qubit}
H = U_J (\chi) + U_C (\hat{Q}),
\end{equation}
where $U_C$ is the Coulomb energy of charge $Q$ on the junction
(which has some finite capacitance $C$). The charge operator
$\hat{Q}$ can be expressed in terms of the phase difference across
the junction as $\hat{Q}=-i\partial_{\chi}$ (see e.g.
\cite{Tinkham96,Zagoskin98}). Due to the presence of the Coulomb
term, phase is no longer a "good" quantum variable, and the
eigenstates of the Hamiltonian (\ref{eq_Hamiltonian_phase_qubit})
become linear combinations of "up" and "down", or "left" and
"right", states (with spontaneous flux $\pm\Phi_0/2$). In other
words, the qubit can now tunnel between the wells of the Josephson
potential, corresponding to certain phases
(Fig.\ref{fig_Ilichev01}). (This description is
appropriate as long as the Coulomb energy does not exceed the
Josephson energy, otherwise the natural starting point would be
the states with definite {\em charge} on the junction; we do not
consider  such systems ({\em charge} qubits) here, but they were
successfully implemented experimentally\cite{Nakamura:1999}.)


Coherent quantum tunneling was indeed observed in an RF
SQUID\cite{Freidman:2000}. Simultaneously, this effect was
obtained in a different system, consisting of a small inductance
loop with {\em three} Josephson junctions (the so called
persistent current qubit, \cite{vdWal:2000}).

The advantage of the latter design is as follows. As we have seen
in Section \ref{sec_1_pi_junction}, the degenerate states appear
in the RF SQUID only if the self-inductance of the loop is large
enough, and they carry comparatively large spontaneous fluxes,
$\pm\Phi_0/2$. Therefore they will couple to the external degrees
of freedom and reduce the time $\tau_d$ during which the system
maintain its quantum coherence. Even more important is the fact
that the resulting potential barrier is comparatively high, which
may prohibit the tunneling we are after. In an experiment
\cite{Freidman:2000} the coherent tunneling was indeed observed
not between the lowest, but between the excited states in the
wells.

In case of the three-junction loop of negligible self-inductance,
the fluxoid quantization condition
(\ref{eq_fluxoid_quantization_2}) leaves two independent Josephson
phase differences in the circuit:
\begin{equation}\label{eq_fluxoid_in_Mooij}
\chi_1+\chi_2+\chi_2+2\pi\frac{\Phi_{\rm external}}{\Phi_0} = 2\pi
n.
\end{equation}
The Josephson energy,
\begin{equation}\label{eq_Mooij_qubit_energy}
U_J=-E_{J1}\cos(\chi_1)-E_{J2}\cos(\chi_2)-E_{J3}\cos(\chi_3)
\end{equation}
 of the system forms a 2D potential profile (e.g. as a
function of $\chi_1$, $\chi_2$), which depends on the external
flux $\Phi_{\rm external}\equiv f\times\Phi_0$ as a parameter.

If $f=0.5$, and the $E_J$'s are comparable, this potential has two
degenerate minima; unlike the case of the RF SQUID, the potential
barrier can be small, as are the spontaneous fluxes corresponding
to the "left" and "right" states.

\subsection{Rationale and proposed designs for qubits with $d$-wave superconductors}

One of the main problems with the designs of the previous section
is the necessity to artificially break the \cal{T}-symmetry of the
system by putting a half flux quantum through it. Estimates show
\cite{Blatter01} that the required relative accuracy is
$10^{-5}-10^{-6}$. The micron-size qubits must be positioned close
enough to each other to make possible their coupling; the
dispersion of qubit parameters means that applied fields must be
locally calibrated; this is a formidable task given such sources
of field fluctuations as fields generated by persistent currents
in qubits themselves, which depend on the state of the qubit;
field creep in the shielding; captured fluxes and magnetic
impurities. Moreover, the circuitry which produces and tunes the
bias fields is an additional source of decoherence in the system.

These problems are avoided if the qubit is \textit{intrinsically
bistable}. The most straightforward way to achieve this is to
substitute the external flux by a static \textit{phase shifter}, a
Josephson junction with
unconventional superconductors with nonzero equilibrium phase shift $\chi_{0}%
$. For example, three-junction (persistent current, Mooij) qubits
would require an extra $\pi$-junction ($\chi_{0}=\pi$)
\cite{Blatter01}.  The only difference compared to the case of an
external magnetic field bias is in the prevalent decoherence
sources: instead of noise from field-generating circuits we will
have to take into account intrinsic decoherence from nodal
quasiparticles and interface bound states (see below).

\begin{figure}
\begin{center}
\includegraphics[scale=0.4]{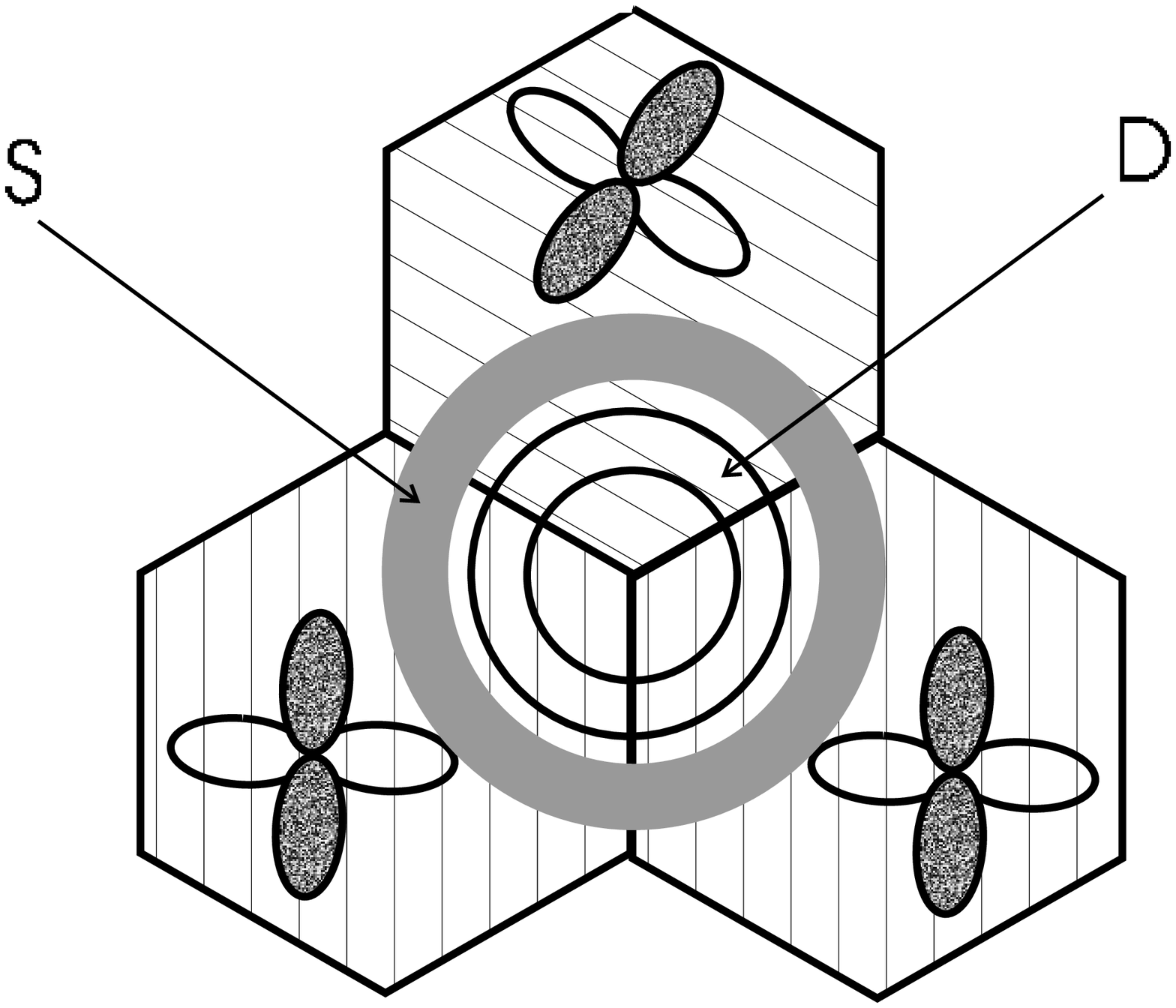}
\end{center}
\caption{A tricrystal high-T$_c$ qubit \cite{Newns02}. The ring D
 formed of high-T$_c$ film contains one $\pi$-junction and therefore
 supports a spontaneous flux
 $\pm\Phi_0/2$\cite{Tsuei94}.
 The ring S of conventional superconductor screens this flux from
 the environment.}\label{fig_Newns}
\end{figure}


Another suggestion \cite{Newns02} is based on the same tricrystal
high-T$_c$ ring geometry as in exp eriments\cite{Tsuei94,Tsuei00}
(Fig\ref{fig_Newns}). The spontaneous flux $\pm\Phi_0$ generated
by such a structure labels the qubit states $|0\rangle$,
$|1\rangle$. Tunnelling between them, necessary for quantum
operations, is made possible
 by applying a magnetic field in the plane of the
system. Indeed, then the states $|0\rangle$, $|1\rangle$ are no
longer the eigenstates of the Hamiltonian.

 The tricrystal  ring D is surrounded by an s-wave
superconductor ring S, aimed at screening the spontaneous flux
from the environment (including other qubits). Indeed, due to the
 fluxoid quantization condition (\ref{eq_fluxoid_quantization}) in
 the ring S, the total flux through it must be an {\em integer}
 multiple of $\Phi_0$, and therefore the states of the rings D and
 S become entangled (e.g. $\alpha|0\rangle_D\bigotimes|1\rangle_S +
 \beta|1\rangle_D\bigotimes|0\rangle_S$).

This entanglement puts forward an interesting problem.  In order
to perform two-qubit operations, as well as initialization and
readout, it is necessary to make the qubits "visible" to each
other and the outside world. To do this, it is suggested in
\cite{Newns02} to locally destroy superconductivity in the
screening ring  by using a superconducting field effect transistor
(SUFET) (not shown) (i.e. by applying a gate voltage to the part
of the screening ring). Will this transition collapse the wave
function of the qubit, destroying quantum coherence between
$|0\rangle_D$ and $|1\rangle_D$?

\begin{figure}[tbp]
\begin{center}
\includegraphics[scale=0.5]{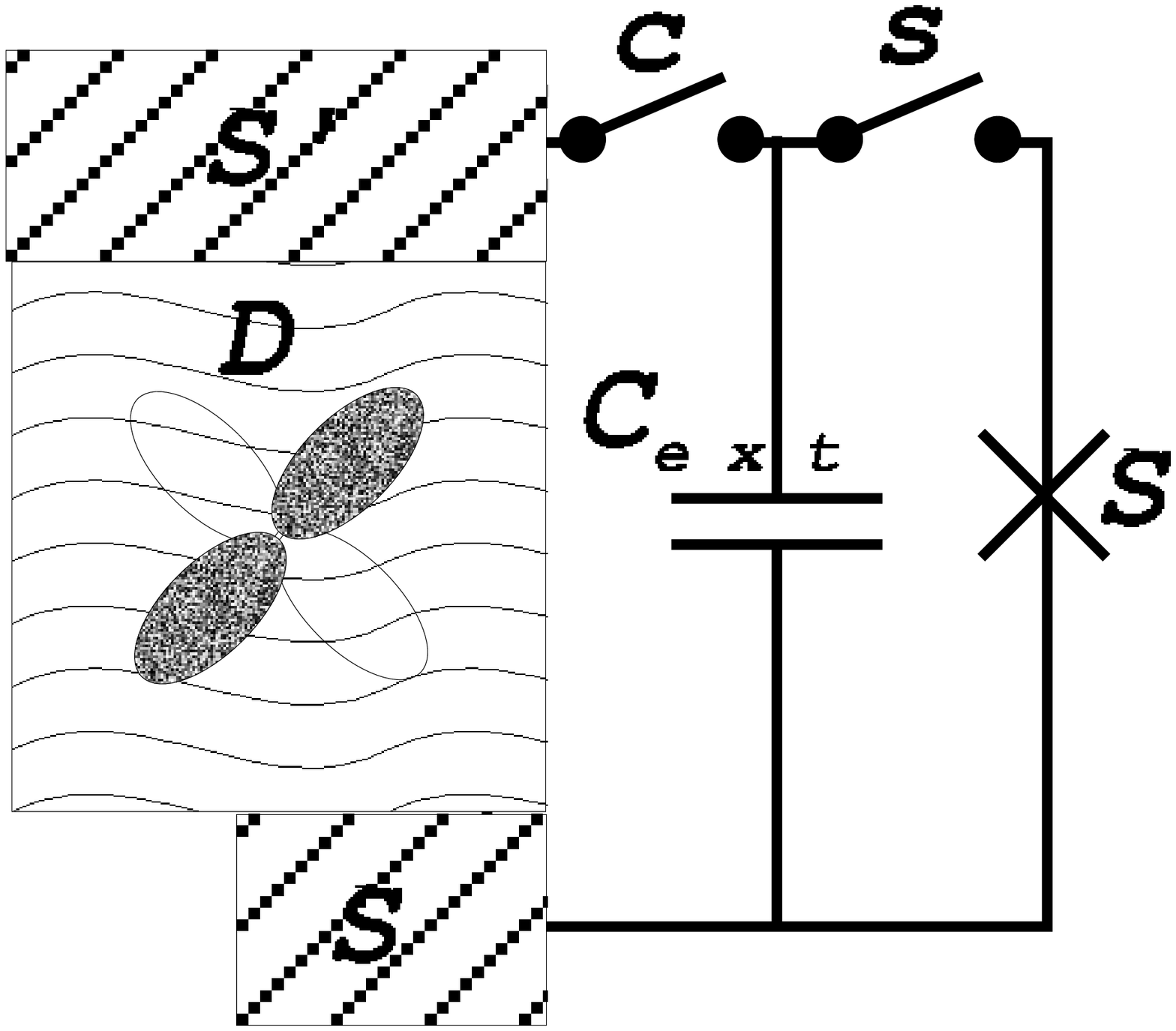}
\end{center}
\caption{"Quiet" SDS' qubit (after \cite{Ioffe99}). Switches $c$
and $s$ connect the SDS' structure to a conventional Josephson
junction, $S$, and a large capacitor, $C_{\rm ext}$.
}\label{fig_Ioffe}
\end{figure}

Now let us consider the case when the bistable $d$-wave system is
employed $dynamically$, that is, when its phase is allowed to
tunnel between the degenerate values. In a so called "quiet" qubit
\cite{Ioffe99} an SDS' junction (effectively two SD junctions in
the (110) direction) put in a small-inductance SQUID loop in
parallel with a conventional Josephson junction and a large
capacitor, Fig.\ref{fig_Ioffe}. One of the SD junctions plays the
role of a $\pi/2$-phase shifter. The other junction's capacitance
$C$ is small enough to make possible tunneling between the $\pi/2$
and $-\pi/2$-states due to the Coulomb energy $Q^{2}/2C$. Two
consecutive SD junctions are effectively a single junction with
equilibrium phases $0$ and $\pi$ (which are chosen as working
states of the qubit, $|0\rangle$ and $|1\rangle$). The proposed
control mechanisms are based on switches $c,~s$. Switch $c$
connects the small S'D junction to a large capacitor, thus
suppressing the tunneling. Connecting $s$ for the duration $\Delta
t$ creates an energy difference $\Delta E$ between $|0\rangle$ and
$|1\rangle$, because in the latter case we have a frustrated SQUID
with $0$- and $\pi$- junctions, which generates $\Phi_{0}/2$
spontaneous flux. This is a generalization of applying a
$\sigma_{z}$ operation to the qubit. Finally, if switch $c$ is
open, the phase of the small junction can tunnel between $0$ and
$\pi$. Entanglement between different qubits is realized by
connecting them through another Josephson junction in a bigger
SQUID loop.(The switches in question are in no way
trivial, since they must operate without destroying quantum
coherence of the system. One possible solution is to use a
frustrated dc-SQUID\cite{Ioffe99}, that is, insert in the wire a
small inductance loop with two equivalent Josephson junctions in
parallel. The total supercurrent through the switch, $I =
I_1(\chi_1) + I_2(\chi_2)$, goes to zero if the phase difference
(tunable by external magnetic flux) $\chi_1-\chi_2=\pi$.
Modifications of this scheme are discussed in \cite{Blatter01}.
Another possibility is to use superconducting single-electron
transistors (SSETs, "parity keys")\cite{Zagoskin99}.)

Due to the absence of currents through the loop during tunneling
between $|0\rangle$ and $|1\rangle$ the authors called it "quiet",
though as we have seen small currents and fluxes are still
generated near SD boundaries.

\begin{figure}[tbp]
\begin{center}
\includegraphics[scale=1]{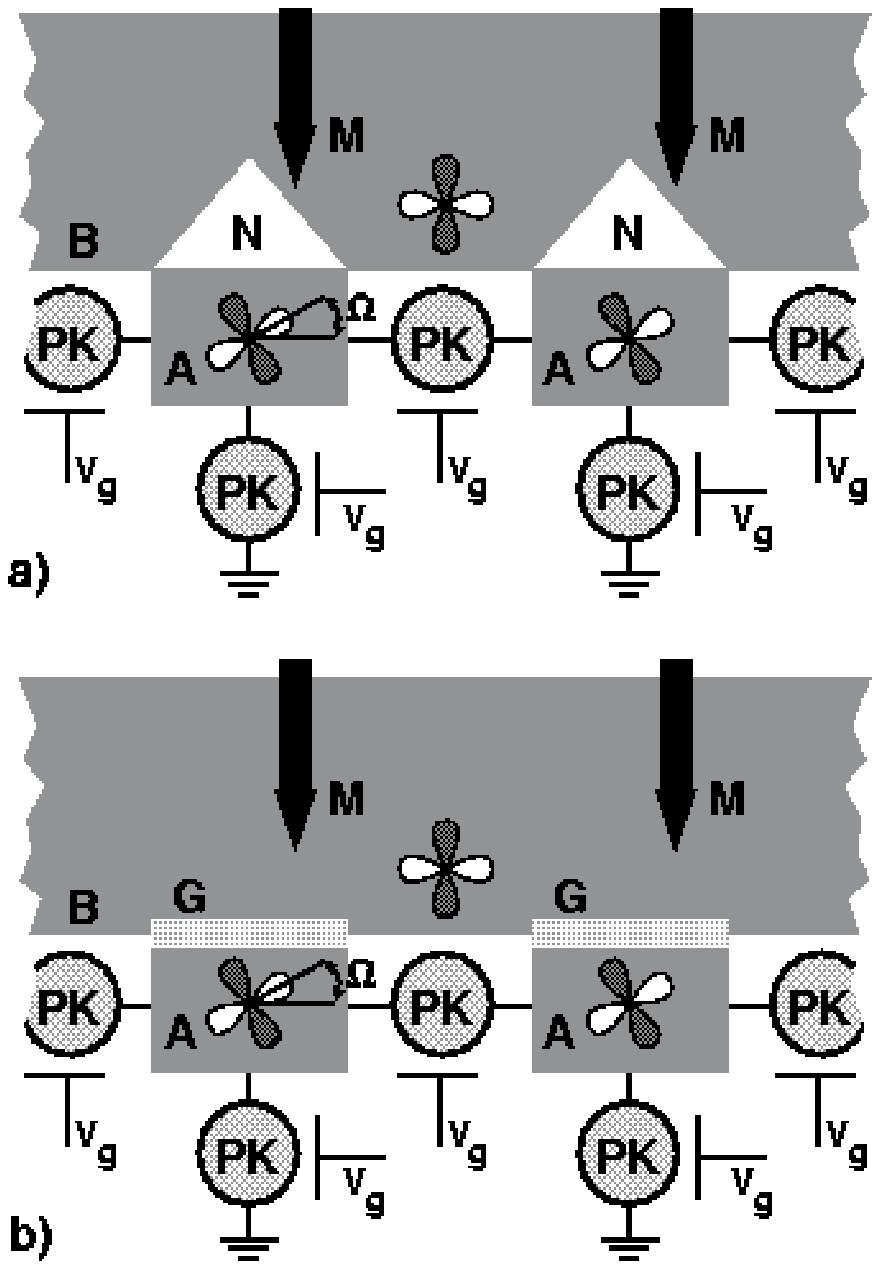}
\end{center}
\caption{A qubit based on an SD (a) or grain boundary DD (b)
Josephson junction\cite{Zagoskin99,Blais00}. Parity keys
(superconducting single electron transistors) are used to connect
the nearest neighbours. The critical current through the parity
key (a small superconducting island) can be tuned by changing the
gate voltage, $V_g$ \cite{Joyez}. $M$ are magnetic force
microscope tips, intended to read out spontaneous fluxes. An
alternative readout scheme relies on direct detection of the phase
on the island, which affects the critical current between the bus
and the ground \cite{Zagoskin02}.   }\label{fig_AZ_DDqubit}
\end{figure}

Another design based on the same bistability
\cite{Zagoskin99,Blais00} only requires one SD or DD boundary
(Fig.\ref{fig_AZ_DDqubit}). Here a small island contacts a massive
superconductor, and the angle between the orientation of the
$d$-wave order parameter and the direction of the boundary can be
arbitrary (as long as it is compatible with bistability
\cite{Amin01,Amin02b}, Fig.\ref{fig_bistability_region}). The
advantage of such a design is, that the potential barrier can be
to a certain extent controlled and
suppressed; moreover, in general the two "working" minima ($-\phi_{0}%
,~\phi_{0}$; the phase of the bulk superconductor across the
boundary is zero) will be separated from each other by a smaller
barrier, than from the equivalent states differing by $2\pi n$.
This allows us to disregard the "leakage" of the qubit state from
the working space spanned by ($|0\rangle$,$|1\rangle$), which
cannot be done in the "quiet" design with the exact
$\pi$-periodicity of the potential profile.

Qubit operations in this system are realized by connecting qubits
to each other and to the ground electrode (normal or
superconducting) through SSETs (or other kind of switches).  When
isolated, a qubit undergoes natural evolution between $|0\rangle$
(state with phase $-\phi_0$) and $|1\rangle$ (phase $\phi_0$),
which realizes  the $\sigma_{x}$ operation. The $\sigma_{z}$
operation
 (that is, adding a controllable phase shift to one of
the states with respect to the other) can be realized by e.g.
connecting the island through a SSET to the massive superconductor
("bus"), the phase of which $\chi\neq0$. The same operation
repeated periodically can be used to block unwanted tunneling (so
called "bang-bang" technique) \cite{Blais00}: physically, if we
keep shifting the levels in the right and left well with respect
to each other, the tunneling becomes suppressed, since they are
practically never in resonance.

\begin{figure}[tbp]
\begin{center}
\includegraphics[scale=0.3]{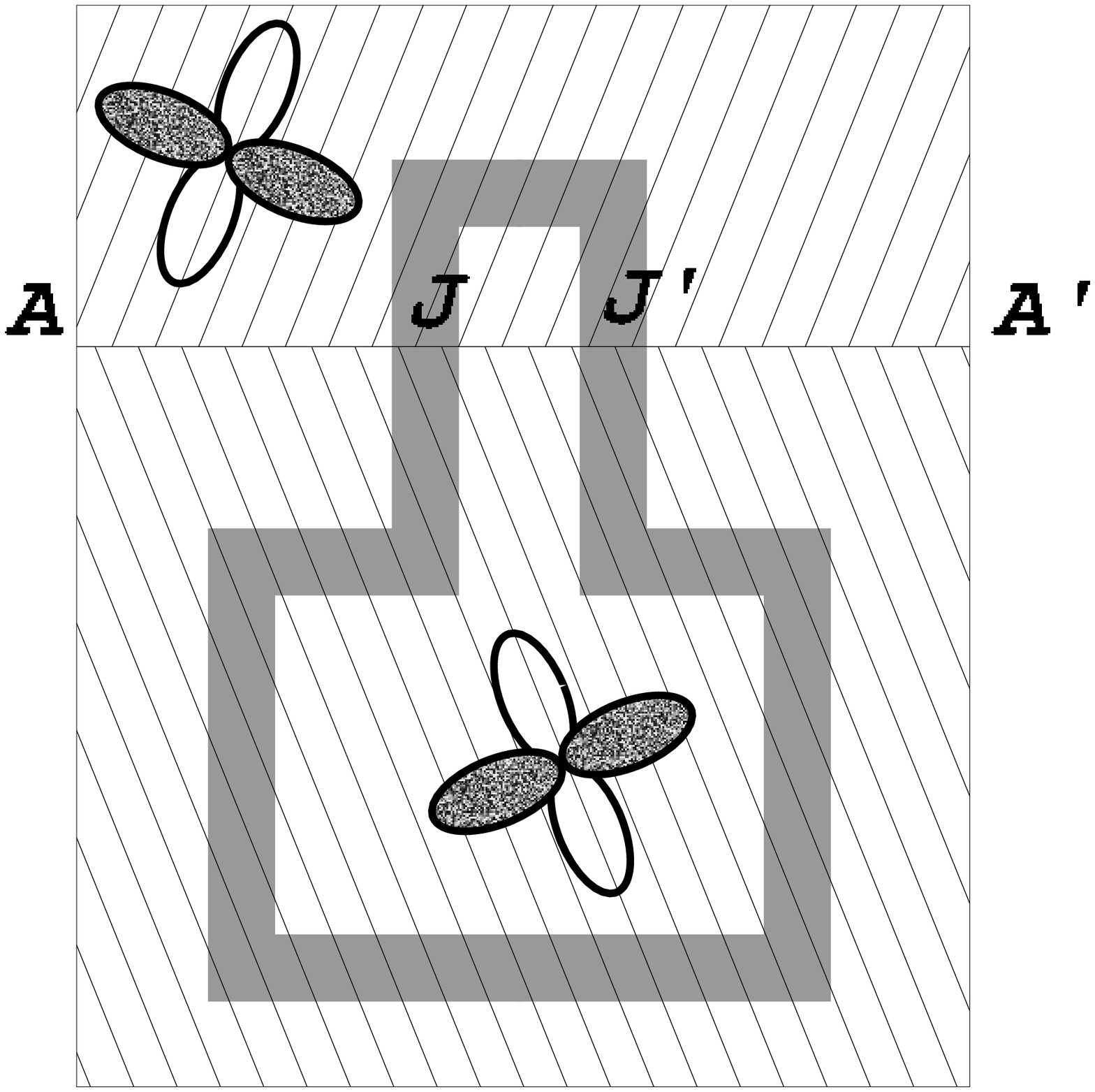}
\end{center}
 \caption{ A silent $d$-wave
qubit\cite{Amin03}. The structure is formed of a high-T$_c$ film
around the grain boundary AA' and contains two grain boundary
junction (J,J'). 
}
\label{fig_silent}
\end{figure}

A better design was suggested in \cite{Amin03} ("silent qubit").
Here two small bistable $d$-wave grain boundary junctions with a
small superconducting island between them are set in a SQUID loop
(Fig. \ref{fig_silent}). As usual, "small" means that the total
capacitance of the system allows phase tunneling: the Coulomb
energy term in (\ref{eq_Hamiltonian_phase_qubit}) is not
negligible.

In the limit of negligible self-inductance of the loop, the
quantization condition (\ref{eq_fluxoid_quantization}) fixes the
sum of the phases to the external flux, $\chi_1+\chi_2 \equiv \phi
= 2\pi\Phi/\Phi_0 $. This leaves only one independent phase
combination, the superconducting phase of the island,
$\theta=(\chi_1-\chi_2)/2$.

\begin{figure}[tbp]
\begin{center}
\includegraphics[scale=0.7]{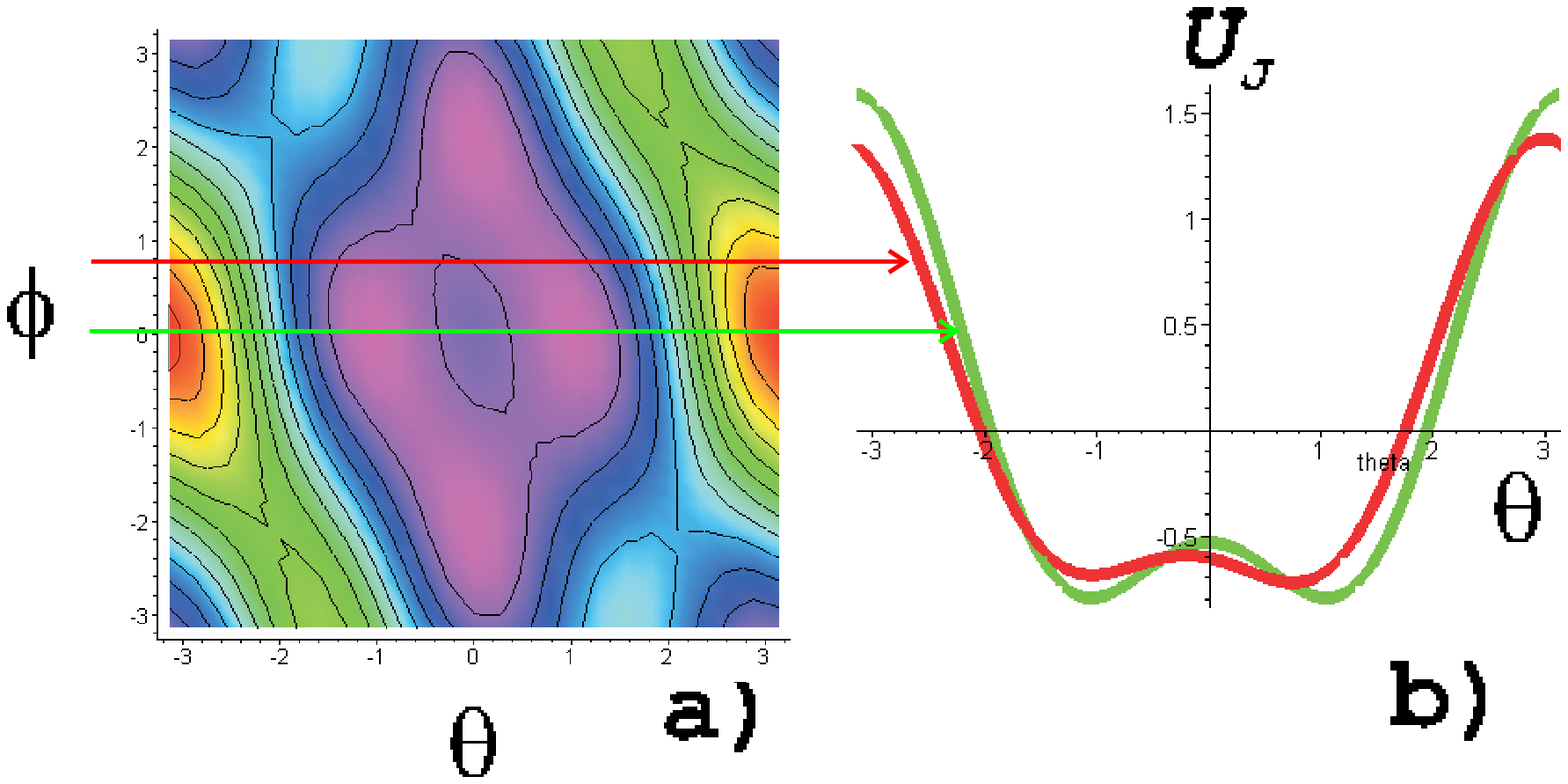}
\end{center}
 \caption{ The Josephson energy profile for the silent
qubit\cite{Amin03} as a function of the island phase, $\theta$;
$\gamma_1=\gamma_2=\pi/4;$ $\eta=0.5$. In the limit of negligible
self-inductance the potential depends on the external magnetic
flux, $\Phi_{\rm ext}=\Phi_0\times (\phi/2\pi) $ as on parameter
(a). In the absence of external flux the potential has two
degenerate minima; degeneracy is lifted by finite external flux,
which also affects the height of the potential barrier(b).}
\label{fig_silent2}
\end{figure}

Keeping for simplicity only the first two harmonics in the
current-phase relation  of the junctions,
\begin{eqnarray}
I_{i} = I_{i}^{(I)}\sin\chi_1-I_{i}^{(II)}\sin 2\chi \equiv
\nonumber\\
\equiv I_{0i} \left[ \sin\chi_{i} \sin \gamma_i - \sin 2\chi_{i}
\cos \gamma_i \right], \label{eq_current_phase_silent}
\end{eqnarray}
  we find for the
Josephson potential of the qubit the expression (Fig.
\ref{fig_silent2})
\begin{eqnarray}\label{eq_silent_qubit_U}
U_J(\theta,\phi)=-(I_{01}/2e)\left[f(\phi/2+\theta,\gamma_1)+ \eta
f(\phi_2-\theta,\gamma_2)\right].
\end{eqnarray}
Here
$f(\varphi,\gamma)=\cos(\varphi)\sin(\gamma)-(1/2)cos(2\varphi)\cos(\gamma)$,
and $\eta=I_{02}/I_{01}$. Parameter $\gamma\in[0,\pi/2]$ provides
a convenient parametrization for the current-phase dependence in a
DD junction.

In the absence of the external flux ($\phi=0$) the qubit potential
$U_J(\theta,0)$ has two degenerate minima. Moreover, if the
junctions only differ in the amplitude of critical current, but
have the same $\gamma$, there is no spontaneous current in the
loop in either minimum, which means that the qubit is decoupled
from the external magnetic fields (if we disregard spontaneous
currents in the junctions themselves, which can be very small
\cite{Amin01,Amin02b,Amin02}). This justifies the moniker
"silent".

Both the barrier and the bias between the wells can be controlled
by the external flux. It is noteworthy that the corrections are of
 at least second order in $\phi$, which drastically reduces the influence of fluctuations
 in the external circuitry\cite{Amin03}. The mechanism of noise
 reduction is similar to that of the "quantronium"
 qubit\cite{Vion}, but the "sweet spot" (an extremal point on the energy
 surface)  appears already on the
 classical level, and at zero external field.

Finally, let us briefly mention two more proposals.

A "no tunneling" design\cite{Amin03a} combines the ideas of CBJJ
(current-based Josephson junction) qubits \cite{Yu02,Martinis02}
and the intrinsic bistability of d-wave junctions. In a single
bistable Josephson junction, the potential barrier between the
degenerate levels corresponding to phase difference $\pm\chi_0$ is
too high to allow tunneling. The transitions between the states
are realized through Rabi transitions via an auxiliary energy
level, situated above the top of the barrier. Rabi transitions are
induced by applying an external high-frequency field.

"Dot/antidot" proposals are based on
 the spontaneous flux (less than a flux quantum) generated in a
high-T$_c$ island or around a hole in bulk high-T$_c$ due to the
presence of a subdominant order parameter\cite{Amin02,Zagoskin02}
(see the end of Section \ref{sec_1_Josephson effect}), but so far
lack experimental support.

\subsection{Fabrication and experiment}

Due to fabrication difficulties, as well as expected problems with
decoherence from e.g. nodal quasiparticles (see Section
\ref{sec_decoherence}), experimental research on $d$-wave qubits
is not as far advanced as on the devices with conventional
superconductors. The experimental confirmation of quantum
behaviour in these systems is still missing. Nevertheless several
recent successes should be noted.

Arrays of half-flux quanta were realized and manipulated
(classically) in YBCO-Nb zigzag junctions \cite{Hilgenkamp03}.
Each facet of the junction effectively constitutes a $\pi$-ring,
which supports spontaneous flux of $\pm\Phi_0$. Interaction
between the fluxes is mainly due to the superconducting
connection, which leads to robust antiferromagnetic ordering. In
its absence, a weaker, magnetic interaction establishes
ferromagnetic flux ordering. The authors consider the possibility
of using their structures in qubit design.

\begin{figure}[tbp]
\begin{center}
\includegraphics[scale=0.7]{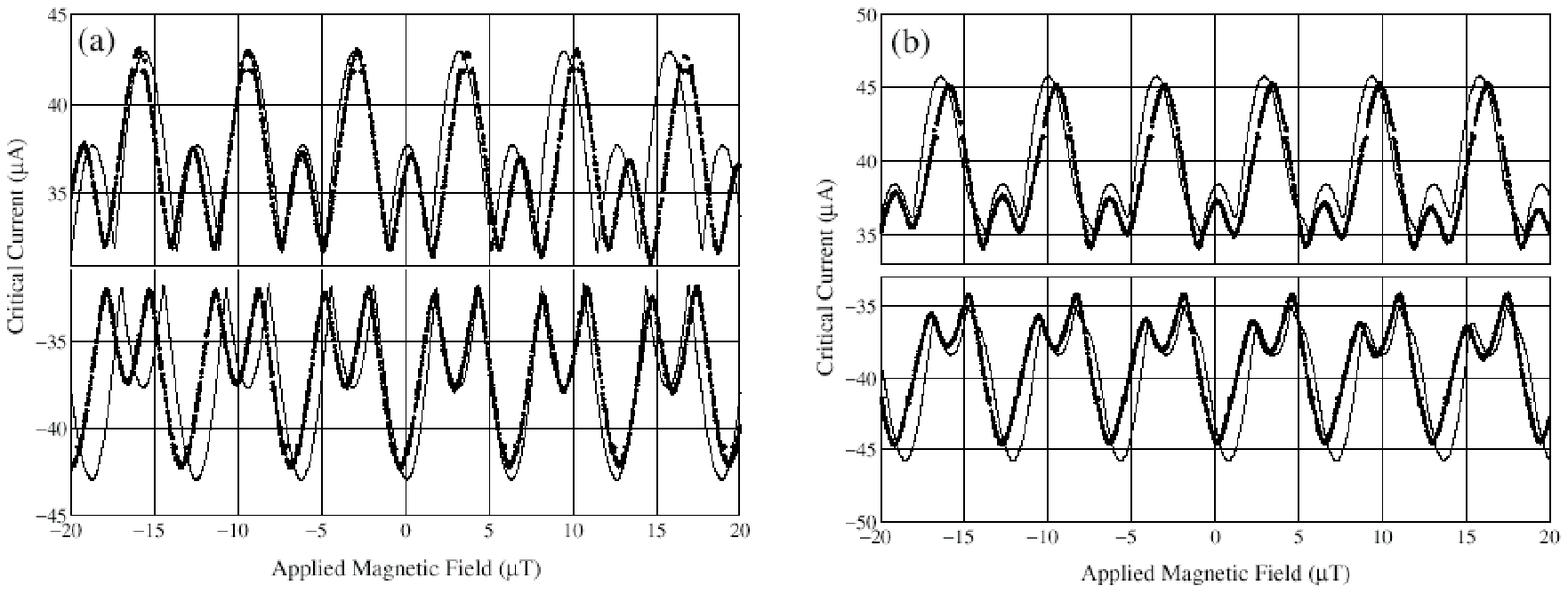}
\end{center}
 \caption{Critical currents of two nominally identical dc SQUIDs,
 based on YBCO grain boundary junctions, as a function of applied magnetic field\cite{Lindstrom03}.
 Dots: experiment. Thin line: theoretical fit.
 Fitting parameters: (a) $I_1^{(I)}=9\:\mu A;$  $I_1^{(II)}=3.7\:\mu A;$
 $I_2^{(I)}=0.3\:\mu A;$ $I_2^{(II)}=22.7\:\mu A;$
 (b) $I_1^{(I)}=7.8\:\mu A;$  $I_1^{(II)}=5.3\:\mu A;$
 $I_2^{(I)}=3\:\mu A;$ $I_2^{(II)}=4.3\:\mu A.$}
\label{fig_dc_SQUID}
\end{figure}

Good quality submicron grain boundary YBCO junctions were
fabricated and bistable energy vs. phase dependence was
demonstrated \cite{Ilichev01,Tzalenchuk03}. Dc SQUIDs YBCO
(15$\times$15 $\mu{\rm m}^2$ square loops with nominally $2\:\mu$m
wide grain boundary junctions) were fabricated and tested, and
their classical behaviour is very well described by the existing
theory \cite{Lindstrom03} (Fig. \ref{fig_dc_SQUID}). Like in
(\ref{eq_current_phase_silent}), only two harmonics of
current-phase dependence were considered. From fitting the
experimental data, we had to conclude that the junctions in the
same SQUID have not only different critical current amplitudes,
but different ratios of first to second harmonic ($\gamma_1 \neq
\gamma_2$), probably due to the variation in the grain boundary
properties over a distance of $\sim$ 15 $\mu$m. This is one reason
for putting the junctions in the silent qubit, Fig.
\ref{fig_silent}, close to each other, as it was done when
fabricating its prototype\cite{Amin03}.

\subsection{Decoherence in $d$-wave qubits}\label{sec_decoherence}

Decoherence is the major concern for any qubit realization,
especially for solid state qubits, due to abundance of low-energy
degrees of freedom. In superconductors, this problem is mitigated
by the exclusion of quasiparticle excitations due to the
superconducting gap. This explains also why the very fact of
existence of gapless excitations in high-T$_c$ superconductors
long served as a deterrent against serious search for macroscopic
quantum coherence in these systems. An additional source of
trouble may be zero-energy states in DD junctions.

Nevertheless, recent theoretical analysis of DD junctions
\cite{Fominov03a,Amin03b,Fominov03}, all using quasiclassical
Eilenberger equations, shows that the detrimental role of nodal
quasiparticles and ZES could be exaggerated.

Before turning to these results, let us first do a simple estimate
of dissipation due to nodal quasiparticles in bulk $d$-wave
superconductors.

Consider, for example, a three-junction ("Mooij") qubit with
d-wave phase shifters. The $|0\rangle$ and $|1\rangle$ states
support, respectively, clockwise and counterclockwise persistent
currents around the loop, with superfluid velocity
$\mathbf{v}_{s}$. Tunnelling between these states leads to nonzero
average $\langle\dot{\mathbf{v}}_{s}^{2}\rangle$ in the bulk of
the superconducting loop.

Time-dependent superfluid velocity produces a local electric field
\begin{equation}
{\mathbf{\
E}}=-\frac{1}{c}\dot{\mathbf{A}}=\frac{m}{e}\dot{\mathbf{v}}_{s},
\end{equation}
and quasiparticle current ${\mathbf{j}}_{qp}=\sigma{\mathbf{E}}.$
The resulting average energy dissipation rate per unit volume is
\begin{equation}
\label{q5}\dot{\mathcal{E}} = \sigma E^{2} \approx
m\tau_{qp}\langle\bar {n}(v_{s})\dot{v}_{s}^{2}\rangle.
\end{equation}
Here $\tau_{qp}$ is the quasiparticle lifetime, and
\begin{equation}
\label{q6}\bar{n}(v_{s}) =
\int_{0}^{\infty}d\epsilon\bar{N}(\epsilon)\left[
n_{F}(\epsilon-p_{F}v_{s})+n_{F}(\epsilon+p_{F}v_{s})\right]
\end{equation}
is the effective quasiparticle density. The angle-averaged density
of states inside the d-wave gap is \cite{Xu95}
\begin{equation}
\label{q7}\bar{N}(\epsilon)\approx
N(0)\frac{2\epsilon}{\mu\Delta_{0}},
\end{equation}
where $\mu=\frac{1}{\Delta_{0}}\frac{d|\Delta(\theta)|}{d\theta}$,
and $\Delta_{0}$ is the maximal value of the superconducting order
parameter. Substituting (\ref{q7}) in (\ref{q6}), we obtain
\begin{equation}
\label{q8}\bar{n}(v_{s}) \approx N(0)\frac{2}{\mu\Delta_{0}}\left(
-T^{2}\right) \left( \mathrm{Li}_{2}\left(
-e^{-\frac{p_{F}v_{s}}{T}}\right) +\mathrm{Li}_{2}\left(
-e^{\frac{p_{F}v_{s}}{T}}\right) \right) ,
\end{equation}
where Li$_{2}(z) = \int_{z}^{0} dt \frac{\ln(1-t)}{t}$ is the
dilogarithm. Expanding for small $p_{F}v_{s}\ll T$, and taking
into account that Li$_{2}(-1)=-\pi^{2}/12$,
Li$_{2}^{\prime}(-1)=\ln2$, and
Li$_{2}^{\prime\prime}(-1)=-1/2+\ln2,$ we obtain
\begin{equation}
\label{q9}\bar{n}(v_{s})\approx\frac{N(0)}{\mu\Delta_{0}}\left(
\frac{\pi ^{2}T^{2}}{3}+ \left( p_{F}v_{s}\right) ^{2}\right) .
\end{equation}
The two terms in parentheses correspond to thermal activation of
quasiparticles and their "Cherenkov" generation by
current-carrying state. Note that finite quasiparticle density
does not lead by itself to any dissipation.

In the opposite limit ($T\ll p_{F}v_{s}$) only the "Cherenkov"
contribution remains,
\begin{equation}
\label{q11}\bar{n}(v_{s})\approx\frac{N(0)}{\mu\Delta_{0}} \left( p_{F}%
v_{s}\right) ^{2}%
\end{equation}
(since Li$_{2}(z)\sim-(1/2)(\ln z)^{2}$ for large negative $z$,
and $\sim z$ for small $z$).

The energy dissipation rate provides the upper limit
$\tau_{\epsilon}$ for the decoherence time (since dissipation is
sufficient, but not necessary condition for decoherence). Denoting
by $I_{c}$ the amplitude of the persistent current in the loop, by
$L$ the inductance of the loop, and by $\Omega$ the effective
volume of d-wave superconductor, where it flows, we can write
\begin{equation}
\label{q10}\tau_{\epsilon}^{-1}=\frac{2\dot{\mathcal{E}}\Omega}{LI_{c}^{2}%
}\approx\frac{2m\tau_{qp}N(0)\Omega\left(
\frac{\pi^{2}T^{2}}{3}\langle\dot
{v}_{s}^{2}\rangle+p_{F}^{2}\langle v_{s}^{2}\dot{v}_{s}^{2}%
\rangle\right) }{\mu\Delta_{0} LI_{c}^{2}}.
\end{equation}
Note that the thermal contribution to $\tau_{\epsilon}^{-1}$ is
independent on the absolute value of the supercurrent in the loop
($\propto v_{s}$), while the other term scales as $I_{c}^{2}$.
Both contributions are proportional to $\Omega$ and (via
$\dot{v}_{s}$) to $\omega_{t}$, the characteristic frequency of
current oscillations (i.e. tunneling rate between clockwise and
counterclockwise current states).

It follows from the above analysis that the intrinsic decoherence
in a $d$-wave superconductor due to nodal quasiparticles can be
minimized by decreasing the amplitude of the supercurrent through
it, and the volume of the material where \textit{time-dependent}
supercurrents flow. From this point of view the designs with phase
shifters, as well as the Newns-Tsuei and "no tunneling" designs
are at a disadvantage (the latter, because the microwave field
necessary to produce Rabi transitions will affect the whole
sample).

Now let us estimate dissipation in a DD junction. First, following
\cite{Zagoskin98a,Zagoskin99}, consider a DND model with ideally
transmissive ND boundaries. Due to tunneling, the phase will
fluctuate, creating a finite voltage on the junction,
$V=(1/2e)\dot{\chi}$, and normal current $I_n=GV$. The
corresponding dissipative function and decay decrement are
\begin{eqnarray}\label{eq_dissipative_function_decay_decrement}
{\cal F}=\frac{1}{2}\dot{\cal E} =
\frac{1}{2}GV^2=\frac{G\dot{\chi}^2}{2}\left(\frac{1}{2e}\right)^2;\\
\gamma = \frac{2}{M_Q\dot{\chi}}\frac{\partial{\cal
F}}{\partial\dot{\chi}} = \frac{G}{4e^2M_Q}=\frac{4
N_{\bot}E_Q}{\pi}.
\end{eqnarray}
Here $E_Q=e^2/2C$, $M_Q=C/16e^2=1/32E_Q$, $N_{\bot}$ are the
Coulomb energy, effective "mass" and number of quantum channels in
the junction respectively. The latter is related to the critical
Josephson current $I_0$ and spacing between Andreev levels in the
normal part of the system $\bar{\epsilon}=v_F/2L$ via
\begin{equation}\label{eq_N_definition}
I_0=N_{\bot}e\bar{\epsilon}.
\end{equation}

  We require, that $ \gamma/\omega_0 \ll 1,$
where $\omega_0=\sqrt{32N_{\bot}E_Q\bar{\epsilon}}/\pi$ is the
frequency of small phase oscillations near a local minimum. This
means,
\begin{equation}\label{eq_no_dissipation}
N_{\bot}\ll \frac{\bar \epsilon}{E_Q}.
\end{equation}
The above condition allows a straightforward physical
interpretation. In the absence of thermal excitations, the only
dissipation mechanism in the normal part of the system is through
the transitions between Andreev levels, induced by fluctuation
voltage. These transitions become possible, if $\bar{\epsilon} <
2e\bar{V} \sim \sqrt{\bar{\dot{\chi}^2}}\sim \omega_0,$ which
brings us back to (\ref{eq_no_dissipation}). Another
interpretation of this criterion arises if we rewrite it as
$\omega_0^{-1} \gg (v_F/L)^{-1}$ \cite{Zagoskin99}. On the
right-hand side we see time for a quasiparticle to traverse the
normal part of the junction. If it exceeds the period of phase
oscillations (on the left-hand side), Andreev levels simply don't
have time to form. Since they provide the only mechanism for
coherent transport through the system, the latter is impossible,
unless our "no dissipation" criterion holds.

For the  thickness of the normal layer $L\sim 1000\:\AA$ and
$v_F\sim 10^7$ cm/s this criterion limits $\omega_0 <
10^{12}\:{\rm s}^{-1}$, which is a comfortable two orders of
magnitude above the usually obtained tunneling splitting in such
qubits ($ \sim 1$ GHz) and can be accommodated in the above
designs. Nevertheless, while presenting a useful qualitative
picture, the DND model is not adequate for the task of extracting
quantitative predictions. For example, the coherence length $l_T$
in the normal metal can be very large, while in the high-T$_c$
compound it is short. Therefore the estimates for crucial
parameters (like $\bar{\epsilon}$) based on the assumption $l_T\gg
L$ can be wrong. Moreover, the assumption of ideal ND boundaries
is not realistic.

A calculation\cite{Newns02}, which used a model of a DD junction
interacting with a bosonic thermal bath, gave an optimistic
estimate for the quality of the tricrystal qubit $Q>10^8$.

The role of size quantization of quasiparticles in small DD and
SND structures was  suggested in \cite{Zagoskin99,Ioffe99}. The
importance of the effect is that it would exponentially suppress
the quasiparticle density and therefore the dissipation below
temperature of the quantization gap, estimated as $1-10$ K.
Recently this problem was investigated\cite{Fominov03} for a
finite width DD junction. Contrary to the expectations, the size
quantization as such turned out to be effectively absent on the
scale exceeding $\xi_0$ (that is, practically irrelevant). On the
other hand, the finite transverse size of the system imposed an
effective band structure. Namely,  due to the direction dependence
of the order parameter, a quasiparticle travelling along the
trajectory, bouncing between the sides of the junction, goes
through a periodic 1D off-diagonal potential. The influence of
this band structure on dissipation in the system is not
straightforward. Ironically, from the practical point of view this
is a moot point, since the decoherence time from the
quasiparticles in the junction, estimated in
\cite{Fominov03,Fominov03a}, already corresponds to the quality
factor $\tau_{\varphi}/\tau_g\sim 10^6$, which exceeds by two
orders of magnitude the theoretical threshold allowing to run a
quantum computer indefinitely.

The expression for the decoherence time obtained in
\cite{Fominov03,Fominov03a},
\begin{equation}\label{eq_Fominov_decoherence}
\tau_{\varphi}=\frac{4e}{\delta\phi_2 I(\Delta_t/e)},
\end{equation}
where $\delta\phi$ is the difference between equilibrium phases in
degenerate minima of the junction (i.e. $\delta\phi=2\chi_0$ in
other notation), contains the expression for quasiparticle current
in the junction at finite voltage $\Delta_t/e$ (where $\Delta_t$
is the tunneling rate between the minima). This agrees with our
back-of-the-envelope analysis: phase tunneling leads to finite
voltage in the system through the second Josephson relation and
with finite voltage comes quasiparticle current and decoherence.
The aforementioned quality factor is defined as
$Q=\tau_{\varphi}\Delta_t/2\hbar$, that is, we compare the
decoherence time with the tunneling time. This is a usual
optimistic estimate, since it will certainly take several
tunneling cycles to perform a quantum gate operation.


It turns out that a much bigger threat is posed by the
contribution from zero energy bound states, which can be at least
two orders of magnitude larger. We can see this qualitatively from
(\ref{eq_Fominov_decoherence}): a large density of quasiparticle
states close to zero energy (i.e., on Fermi level) means that even
small voltages create large quasiparticle currents, which sit in
the denominator of the expression for $\tau_{\varphi}.$
Fortunately, this contribution is suppressed in the case of ZES
splitting, and such splitting is always present due to, e.g.,
finite equilibrium phase difference across the
junction\cite{Fominov03}.

A similar picture follows from the analysis presented in
\cite{Amin03b}. A specific question addressed there is especially
important: it is known that $RC$-constant measured in DD junctions
is consistently 1 ps over a wide range of junction sizes
\cite{Tzalenchuk03}, and it is tempting to accept this value as
the dissipation rate in the system. It would be a death knell for
any quantum computing application of high-T$_c$ structures, and
nearly that for any hope to see there some quantum effects.
Nevertheless, it is not quite that bad. Indeed, we saw that ZES
play a major role in dissipation in a DD junction, but are
sensitive to phase differences across it\cite{FN1}. Measurements of the $RC$
constant are done in the resistive regime, when a finite voltage
exists across the junction, so that the phase difference grows
monotonously in time, forcing ZES to approach the Fermi surface
repeatedly. Therefore $\tau_{RC}$ reflects some averaged
dissipation rate. On the other hand, in a free junction with not
too high a tunneling rate phase differences obviously tend to
oscillate around $\chi_0$ or $-\chi_0$, its equilibrium values,
and do not spend much time near zero or $\pi$; therefore ZES are
usually shifted from the Fermi level, and their contribution to
dissipation is suppressed.



\begin{figure}[tbp]
\begin{center}
\includegraphics[scale=0.6]{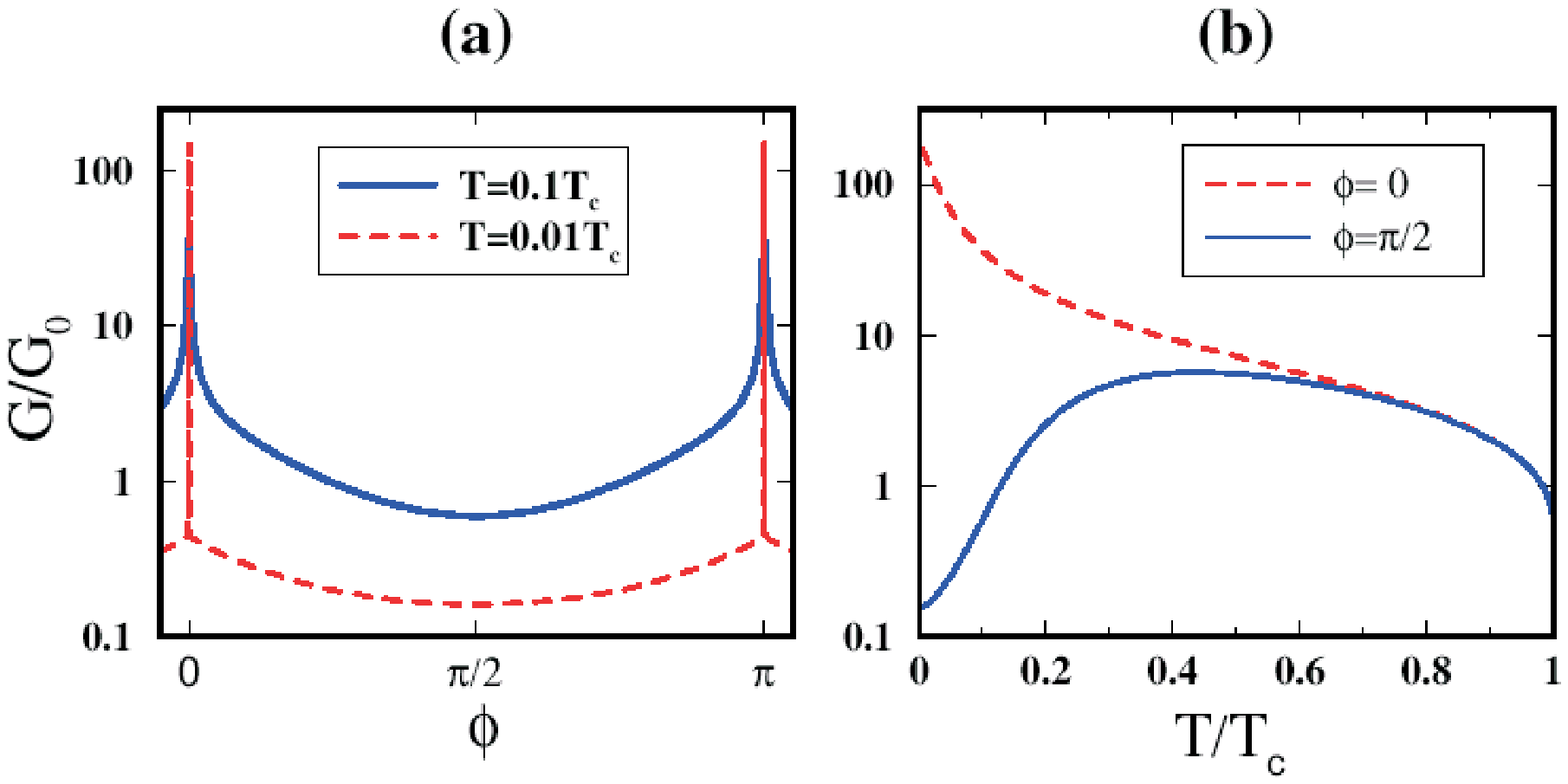}
\end{center}
 \caption{Phase (a) and temperature (b) dependence of the normal conductance of a
 DD junction \cite{Amin03b}.}
\label{fig_AminSmirnov}
\end{figure}

This qualitative picture is confirmed by a detailed calculation\cite{Amin03b}
(Fig. \ref{fig_AminSmirnov}). The decoherence time is related to
the phase-dependent conductance via
\begin{equation}\label{eq_AminSmirnov}
\tau_{\varphi} = \frac{1}{\alpha F(\chi_0)^2 \delta E} \tanh
\frac{\delta E}{2T}.
\end{equation}
Here $\alpha$ is the dissipation coefficient, $\delta E$ is
interlevel spacing in the well, and
\begin{equation}
G(\chi)=4e^2\alpha \left(\partial_{\chi}F(\chi)\right)^2.
\end{equation}
For a realistic choice of parameters Eq.(\ref{eq_AminSmirnov})
gives a conservative estimate $\tau_{\varphi}=1-100$ ns, and
quality factor $Q \sim 1-100$. This is, of course, too little for
quantum computing, but quite enough for observation of quantum
tunneling and coherence in such junctions.

\section{Conclusions}

Since the $d$-wave character of superconducting pairing in
high-T$_c$ cuprates  was established, our understanding of
high-T$_c$ Josephson structures and ability to fabricate them
progressed significantly. Now it enables us to ask the question in
the title of this paper: Can high-T$_c$ cuprates  play  a role in
quantum computing?- and to tentatively answer: Yes.

We have seen that  submicron junctions of sufficiently high
quality are now fabricated. Several designs, which take advantage
of the intrinsic bistability of $d$-wave structures, were
developed, fabricated, and tested in classical regime. If the
decoherence time turns out to be large enough, their better
scalability can be a decisive advantage for quantum computing
applications.

This is, of course, a big if. Still, theoretical estimates tend to
be rather optimistic. Even though they widely differ,  all of them
predict $\tau_{\varphi}$ long enough to encourage experimental
search for quantum coherence. There are several compelling reasons
to do that. First, it would be a spectacular result. Second, it
would clarify   why different models give different answers.
Third, it could indeed lead to practical application of high-T$_c$
devices in quantum computing.

\section*{P.S.}
Since this paper was published (2003), macroscopic quantum  tunneling was observed in YBCO \cite{MQT_YBCO} and BiSCCO \cite{MQT_BiSCCO} junctions. This made the "big if" of the previous paragraph somewhat smaller.


\section*{Acknowledgements}

I am grateful to M.H.S. Amin, A. Blais, A. Golubov, E. Il'ichev,
A.N. Omelyanchouk, A. Smirnov, and A. Tzalenchuk for many
illuminating discussions, and to B. Wilson for the careful reading
and helpful comments on the manuscript.



\begin{thebibliography}{99}
\bibitem{FN1}{A model Hamiltonian for such a system can be written as $H = \frac{\hat{Q}^2}{2C} + U(\chi) + \frac{1}{2}\left[\hat{Q},V(\chi)\right]_+ \sum_{\alpha} c_{\alpha}x_{\alpha} + \sum_{\alpha}\left[ \frac{\hat{p}_{\alpha}^2}{2m_{\alpha}} + \frac{m_{\alpha}\omega_{\alpha}^2x_{\alpha}^2}{2} \right]$. Here $U(\chi)$ is a two-well potential, and $V(\chi)$ is centered near $\chi=0$. }
\end{thebibliography}
\end{document}